Genomic variation in a widespread Neotropical bird (*Xenops minutus*) reveals divergence, population expansion, and gene flow

Michael G. Harvey[1,2] and Robb T. Brumfield[1,2]

[1] *Department of Biological Sciences, Louisiana State University, Baton Rouge, LA 70803*

[2] *Museum of Natural Science, Louisiana State University, Baton Rouge, LA 70803*

Contact Information:

Michael G. Harvey (corresponding author)

Telephone: (225) 578-5393

Email: mharve9@lsu.edu

Robb T. Brumfield

Telephone: (225) 202-8892

Email: robb@lsu.edu

Words: 7,087 (including abstract), Tables: 3 (+6 Supplemental), Figures: 4 (+3 Supplemental),

Data: to be archived in Dryad

Abstract

Elucidating the demographic and phylogeographic histories of species provides insight into the processes responsible for generating biological diversity, and genomic datasets are now permitting the estimation of histories and demographic parameters with unprecedented accuracy. We used a genomic single nucleotide polymorphism (SNP) dataset generated using a RAD-Seq method to investigate the historical demography and phylogeography of a widespread lowland Neotropical bird (*Xenops minutus*). As expected, we found that prominent landscape features that act as dispersal barriers, such as Amazonian rivers and the Andes Mountains, are associated with the deepest phylogeographic breaks, and also that isolation by distance is limited in areas between these barriers. In addition, we inferred positive population growth for most populations and detected evidence of historical gene flow between populations that are now physically isolated. Even with genomic estimates of historical demographic parameters, we found the prominent diversification hypotheses to be untestable. We conclude that investigations into the multifarious processes shaping species histories, aided by genomic datasets, will provide greater resolution of diversification in the Neotropics, but that future efforts should focus on understanding the processes shaping the histories of lineages rather than trying to reconcile these histories with landscape and climatic events in Earth history.

Introduction

Lowland humid forests in the Neotropics contain some of the highest biodiversity on the planet (Pearson 1977). A number of hypotheses have been proposed to explain the origins of this diversity, most of which link biological diversification directly to tumultuous landscape changes that led to speciation via the geographic isolation of populations (Moritz et al. 2000; Antonelli et al. 2010). The hypotheses differ in the events and features implicated. These include the origins of major rivers in the Amazon basin (Sick 1967; Capparella 1987; Ribas et al. 2012), uplift of the Andes and other mountain ranges (Chapman 1917, 1926), past fragmentation of humid forest due to expansion of arid habitats (Haffer 1969) or marine transgressions (Nores 1999; Aleixo 2004), edaphic or climatic conditions associated with geologic arches (Lougheed et al. 1999; Wesselingh and Salo 2006), and areas of displacement due to invasion by temperate taxa during colder periods (Erwin 1979; Bush 1994).

Studies evaluating these hypotheses have typically addressed them using gene genealogies to infer the timing of divergence and the geographic location of vicariance. Using the conceptual framework of vicariance biogeography, researchers have searched for shared phylogeographic (or phylogenetic) relationships among taxa that would suggest a common mechanism of biological diversification (e.g., Cracraft and Prum 1988; Brumfield and Capparella 1996; Hall and Harvey 2002; Quijada-Mascareñas et al. 2007). In addition, molecular dating methods have been used to estimate the timing of population divergence events and to compare these dates to hypothesized events in the landscape evolution of the Neotropics (Patton et al. 2000; Weir 2006; Santos et al. 2009; Ribas et al. 2012). Although some general patterns have emerged from these studies, such as the importance of landscape features in delimiting populations and the absence of an increase in diversification during the Pleistocene, no single dominant model relating historical diversification to landscape history has emerged from decades of genetic studies (reviewed in Haffer 1997; Antonelli et al. 2010; Leite and Rogers 2013).

Interrogating processes beyond divergence may prove to be more fruitful in informing species histories (Takahata et al. 1995, Kuhner et al. 2009). For example, signatures of population size changes found in studies of Neotropical organisms (Aleixo 2004; Cheviron et al. 2005; Solomon et al. 2008; D'Horta et al. 2011) may evidence historical increases or decreases in habitat availability. Evidence of gene flow between populations, which may reveal instances of past connectivity between habitats or regions, has been uncovered in a few studies (Patton et al. 1994; Noonan and Gaucher 2005; Maldonado-Coelho et al. 2013). In addition, a few studies have detected the effects of natural selection and sexual selection among populations (Mallet 1993, Turner et al. 2004), which may be linked to past climatic changes or other events. Reconstructing how these diverse processes influenced modern phylogeographic patterns is challenging, but could provide new insight into the history of Neotropical diversification.

The availability of genome-scale datasets is improving inferences concerning the historical diversification of organisms (Li and Durbin 2011, Frantz et al. 2013). Genomic data, when combined with methods that account for coalescent stochasticity, allow for the integration of information across many loci (Edwards and Beerli 2000), and provide greater statistical power for testing models of population history (Pool et al. 2010). Analyses based on genomic data result in narrower confidence intervals in estimates of important parameters such as divergence times, effective population sizes, and migration rates between populations (Smith et al. 2014). Dense sampling across the genome increases the probability of obtaining data from migrant alleles or genomic regions that have been influenced by selection (Carlson et al. 2005). The application of genomic data to Neotropical systems (e.g., Nadeau et al. 2013) promises to allow further investigation of processes important in Neotropical species histories.

Here, we use dense sampling and genome-scale genotyping-by-sequencing (GBS) data to 1) characterize the geographic structure of genetic variation in a widespread lowland Neotropical bird species (*Xenops minutus*; Aves, Furnariidae) and 2) evaluate a series of predictions concerning its historical demography. *X. minutus* is relatively abundant in humid lowland forests west of the Andes from Mexico to northwestern South America, and, east of the Andes, in the Amazon Basin and Atlantic Forest of eastern South America (Remsen 2003). The species exhibits geographic variation in plumage and voice, with this variation classified into 10 parapatrically or allopatrically distributed subspecies (Dickinson 2003; Remsen 2003). Previous phylogeographic studies (Burney 2009; Smith et al. 2014) of *X. minutus* had limited genomic or geographic sampling, but found evidence for geographically isolated mitochondrial clades and deep genome-wide divergence between populations from either side of the Andes, respectively. Our goals were to determine how the population history of *X. minutus* influences modern patterns of genetic diversity, and to attempt to relate this history to the general landscape history of the Neotropics.

Methods

Genetic Data Collection and Processing

We sampled eight vouchered *X. minutus* from each of nine biogeographic areas for a total of 72 individuals (Fig. 1, Table S1). This sample included 7 of the 10 currently recognized subspecies (Dickinson 2003; Remsen 2003). The remaining three subspecies, distributed in Colombia and the northwestern Amazon Basin, were not included because we lacked genetic material. We extracted total DNA from frozen or alcohol-preserved pectoral muscle tissue using a DNeasy tissue extraction kit (Qiagen, Valencia, CA).

We sent 0.3-3.0 □g of each sample to the Cornell Institute of Genomic Diversity for genotyping-by-sequencing (GBS). GBS is a streamlined workflow for generating reduced representation libraries for Illumina sequencing, similar to other forms of RAD-Seq (Baird et al. 2008, Hohenlohe et al. 2010). Details of the laboratory methods can be found in Elshire et al. (2011). In brief, DNA from each sample was digested using the restriction enzyme PstI (CTGCAG), and both a sample-specific barcoded adapter and a common adapter were ligated to the sticky ends of fragments. Samples were pooled and fragment libraries cleaned using a QIAquick PCR purification kit (Qiagen). Libraries were amplified using an 18-cycle PCR with long primers complementary to the barcoded and common adapters, purified again using QIAquick, and quantified using a PicoGreen assay (Molecular Probes, Carlsbad, CA, USA). Samples were run on a partial lane (72 out of 96 samples) of a 100-bp single-end Illumina HiSeq 2000 run at the Cornell Core Laboratories Center.

The Cornell Institute of Genomic Diversity processed raw sequence reads using the UNEAK pipeline, an extension to TASSEL 3.0 (Bradbury et al. 2007). Briefly, UNEAK retains all reads with a barcode, cut site, and no missing data in the first 64 bp after the barcode. Reads are clustered into tags by 100% identity, tags are aligned pairwise, and any tag pairs differing by one bp are called as potential SNPs. To remove sequencing errors, any alleles represented by fewer than five reads or a frequency of less than 5% are filtered out (Table S2). Following processing with the UNEAK pipeline, we collapsed reverse complement tag-pairs and re-called genotypes using the method of Lynch (2009) as implemented in custom perl scripts obtained from T. A. White (White et al. 2013) and available at https://github.com/mgharvey/GBS_process_Tom_White/v1. We removed potential paralogs by filtering out SNPs with heterozygosity greater than 0.75, and we removed SNPs for which genotype calls were missing from more than 20% of the individuals. The hypothetical genomic distribution of the remaining SNP loci was investigated by aligning their tag-pair consensus sequences (with "N" inserted at the SNP site) to the Zebra Finch (*Taeniopygia guttata*) genome (Warren et al. 2010) using blastn (Altschul et al. 1990). *Taeniopygia guttata* is the most closely related species to *X. minutus* with a publicly available genome assembly, although the evolutionary distance between the two is considerable (Hackett et al. 2008). We used custom python scripts (available at http://github.com/mgharvey/misc_Python) to generate input files for further analysis.

Data analysis: Effects of distance and barriers

Isolation by distance and dispersal barriers are known to geographically structure genetic variation in Neotropical birds (Brawn et al. 1996; Cheviron et al. 2005; Cabanne et al. 2007). We evaluated the importance of these isolating forces using Mantel and partial Mantel tests, as well as a Bayesian model-based method. We used the kinship coefficient (Loiselle et al. 1995) calculated in the program SPAGeDi (Hardy and Vekemans 2002) as an index of pairwise genetic relatedness between individuals. The kinship coefficient $F_{ij}$ is the probability that two homologous genes are identical by descent, and is calculated as $F_{ij} = (Q_{ij}-Q_m)/(1-Q_m)$ where $Q_{ij}$ is the probability of identity by state between two individuals of interest for random genes and $Q_m$ is the average probability of identity by state for genes coming from random individuals in the population. $F_{ij}$ is a relatively unbiased estimator with low sampling variance (Hardy and Vekemans 2002).

We tested for isolation by distance across all individuals using a Mantel test comparing $F_{ij}$ and geographic distance between individuals. Geographic distances were calculated as the Euclidean distances between sampling localities in SPAGeDi. To distinguish isolation by distance from discrete genetic breaks we conducted separate Mantel tests within each biogeographic area bounded by a major dispersal barrier, including the Isthmus of Panama, the Andes Mountains, major Amazonian rivers, and the cerrado belt of eastern Brazil that isolates Amazonia from the Atlantic Forest (based on Cracraft 1985, Fig. 1). To investigate isolation due to the dispersal barriers, we used a partial Mantel test that controlled for geographic distance in testing the correlation between $F_{ij}$ and whether individuals were on the same or different sides of putative barriers. We conducted separate analyses including all barriers and for each barrier individually. Only those individuals in the areas adjoining each barrier were used for the barrier-specific tests to remove confounding influences from other barriers. All Mantel and partial Mantel tests were carried out in the R package ecodist (Goslee and Urban 2007) using 10,000 permutations of geographic locations with individuals to determine significance and a jackknifing procedure to estimate standard errors.

Because Mantel and partial Mantel tests assume linear relationships between variables (Legendre and Fortin 2010), are confounded by spatial autocorrelation (Guillot and Rousset 2013), and are unable to directly quantify the relative importance of predictor variables (Bradburd 2013), we also used a new method, BEDASSLE (Bradburd 2013). BEDASSLE overcomes these issues by modeling the covariance in allele frequencies between populations as a function of the predictor variables, and estimating model parameters in a Bayesian framework using a Markov chain Monte Carlo algorithm. We used BEDASSLE to estimate the relative importance of geographic distance and barriers across the entire distribution of *X. minutus*, as well as between each pair of adjacent populations separated by a specific dispersal barrier. We ran BEDASSLE using the beta-binomial model to account for over-dispersion due to variation in demographic histories across populations. All analyses were run for 10 million generations, sampling every 100. We examined traces, marginal and joint marginal parameter distributions, and MCMC acceptance rates every one to five million generations and adjusted tuning parameters according to the suggestions of Bradburd et al. (2013).

Data analysis: Population assignment and admixture

We estimated the number of populations and conducted population assignment of individuals from all SNPs using methods implemented in STRUCTURE 2.3.4 (Pritchard et al. 2000) and Structurama (Huelsenbeck et al. 2011). Given a fixed number of populations (K), STRUCTURE assigns individuals to populations probabilistically such that Hardy-Weinberg equilibrium and linkage equilibrium within populations are maximized. In addition to population assignment, STRUCTURE can be used to identify admixed individuals. We used STRUCTURE without specifying prior information on population membership, and used options for correlated allele frequencies and genetic admixture across populations (Falush et al. 2003). We conducted runs of 1,000,000 generations (after a 10,000-generation burnin) for each value between K=1 and K=15 and calculated Pr(X|K) to assess the results (Pritchard et al. 2000).

Structurama offers the option of jointly estimating the number of populations (K) and the assignment of individuals to populations using a Dirichlet process prior. We treated K as a random variable and provided an exponential distribution with a mean of nine as a prior for K, consistent with the number of biogeographic regions from which individuals were sampled. We also treated both K and the clustering variable α as random variables and examined the influence of three different gamma priors for α: (1,1), (5,1), and (10,1). For each analysis, we ran MCMC chains for 100 million generations, sampling every 25,000, and discarded 25% of the samples as burnin.

To uncover finer scale population structure we used ChromoPainter and fineSTRUCTURE (Lawson et al. 2012) with the subset of SNPs having no missing data across all 72 individuals. ChromoPainter considers each individual a possible recipient of "chunks" of DNA from a panel of donor individuals. It assembles a "coancestry matrix" recording the number of recombination events between each donor and recipient. In our case, we considered all individuals as potential recipients and donors. Although using linked sites provides more power for population inference using this method, we lacked linkage information for our SNPs, so we treated them as unlinked. fineSTRUCTURE then performs model-based clustering using the information in the coancestry matrix. The normalization parameter $c$, or the effective number of "chunks", is used to rescale the elements of the coancestry matrix before calculating the likelihood, and can influence the amount of inferred population structure. We used a $c$ value of 1/(n-1) where n is the sample size, following the recommendation in Lawson et al. (2012) for unlinked data, but also examined the effects of higher and lower $c$ values.

Population structure is sometimes inferred incorrectly due to the presence of isolation by distance (Meirmans 2012). We examined this possibility by conducting partial Mantel tests of the association between $F_{ij}$ and both the set of populations estimated in fineSTRUCTURE and the set of populations estimated from STRUCTURE with K=5 and Structurama with the gamma prior for alpha equal to (1,5), while controlling for geographic distance. Hereafter we refer to these as the fineSTRUCTURE populations and the STRUCTURE/Structurama populations, respectively.

Data analysis: Population expansion and migration

We estimated expansion within and migration between both the fineSTRUCTURE and STRUCTURE/Structurama populations using coalescent modeling in the program LAMARC (Kuhner 2006, 2009). LAMARC has the advantage of being able to jointly estimate population growth and migration, both of which may be important processes influencing genetic variation in populations of tropical taxa (Moritz et al. 2000). We estimated the standardized population mutation rate ($\theta = 4N_e\mu$) and population growth rate (g, where $\theta_t = \theta_{present}^{-gt}$) for each population as well as the migration rate (M = $m/m\mu$, where $m$ is the immigration rate per generation and $m\mu$ is the neutral mutation rate per site per generation) between adjacent populations separated by the dispersal barriers described above. We used the parameter-poor F84 model of sequence evolution because it is much faster than the alternative GTR model in LAMARC and because a simple model should be sufficient given that mutations are infrequent at the loci examined (SNPs represent a single variable site within an ~64 bp alignment). We set the transition/transversion ratio to 2. We used a Bayesian MCMC approach, and placed uniform priors on $\theta$ (log($1\times10^{-6}$, 10)), M (log($1\times10^{-10}$, 100)), and g (linear(-500, 1000)). We conducted 10 initial chains with 1,000 iterations of burnin followed by 10,000 iterations, followed by 2 independent final chains of 5,000 iterations of burnin followed by 10,000,000 iterations. We checked for convergence within and between chains using Tracer v.1.5 (Rambaut and Drummond 2007), and we report estimates from the second final chain.

Data analysis: Natural selection

We conducted a preliminary examination of selection in *X. minutus* using a multi-population outlier scanning approach implemented in BayeScan 2.01 (Foll and Gaggiotti 2008). BayeScan examines $F_{st}$ values between each population and a common migrant gene pool for each locus. $F_{st}$ coefficients are decomposed into a component shared by all loci ($\beta$) and a locus-specific component ($\alpha$) that reflects selection. BayeScan then compares models in which selection ($\alpha$) is and is not incorporated, and estimates the posterior probability for each model at each locus using a reversible-jump Markov chain Monte Carlo (RJ-MCMC) method. The posterior odds, or ratio of posterior probabilities, are used to decide on the best model and to define thresholds to determine sets of outlier markers. BayeScan is robust to complex demographic scenarios that might influence neutral differentiation (Foll and Gaggiotti 2008). We examined the influence of selection based on analyses using both the STRUCTURE/Structurama and fineSTRUCTURE populations. We ran analyses using 20 pilot runs of 5000 iterations, a burn-in of 50,000 iterations, and a final run of 50,000 iterations. Prior odds for the neutral model were set to 10.

Data analysis: Species tree

We estimated the branching structure of populations using a species tree approach for both the fineSTRUCTURE and STRUCTURE/Structurama populations. Species trees were estimated using the coalescent method implemented in SNAPP (Bryant et al. 2012). SNAPP computes the likelihood of a species tree from unlinked biallelic markers rather than explicitly sampling gene trees. Any SNPs missing genotypes from all individuals in any of the populations were removed from the dataset. Also, due to the computational demands of analyzing the full dataset, we reduced each population to two randomly selected individuals (four haplotypes). We used a diffuse gamma prior for $\theta$ ($\alpha = 10$, $\beta = 100$) and a pure birth (Yule) prior for the species tree, with birth rate ($\lambda$) equal to 0.00765. For each population set, we conducted two runs of 5 million generations, sampling every 1,000 generations. We determined the burnin and assessed MCMC convergence by examining ESS values and likelihood plots in Tracer v.1.5 (Rambaut and Drummond 2007). We combined runs and used TreeAnnotator (Rambaut and Drummond 2008) to determine the Maximum Clade Credibility tree and posterior probability values.

Results

Sequencing and datasets

GBS produced a total of 106,784 biallelic SNPs (Table S3). After collapsing reverse complements and filtering for observed heterozygosity and amount of missing data, the final data matrix contained 3,379 SNPs and was 91.1% complete. Data have been deposited in Dryad (submission pending). We recovered hits to the *T. guttata* genome using blastn for 3,247 of these SNPs. Hits were distributed across 31 of the 36 chromosomes, including the Z chromosome (Table S4). The chromosomes without hits were microchromosomes 16, LGE22, LG2, LG5, and MT. The number of hits per chromosome was positively associated with chromosome size ($r^2$ = 0.836, p < 0.001). We note, however, that the short length of GBS loci may result in low mapping accuracy and that *T. guttata* and *X. minutus* are distant relatives and synteny between the two genomes may be low.

Effect of distance and barriers on genetic divergence

Plotting pairwise kinship coefficients between samples relative to geographic distance revealed wide variation in kinship across the distribution of *Xenops minutus* (Fig. 2). Mantel tests showed a signal of isolation by distance based on correlations between the kinship coefficient and geographic distance (Mantel r statistic = -0.4964, p = 0.0001). However, the signal for isolation by distance was less prevalent within areas; only the Napo, Rondônia, and Atlantic Forest areas showed significant (p < 0.01) evidence of isolation by distance and the slopes were generally shallow (Table S5). Partial Mantel tests across all areas and individuals revealed a relationship between kinship and whether individuals were on the same or opposite sides of barriers after controlling for isolation by distance (r = -0.6467, p = 0.0001). Examining each dispersal barrier separately, we found that all nine barriers showed a significant relationship (p < 0.01) with the kinship coefficient, and the slope of the Mantel correlation was generally steeper than in the within-area isolation by distance comparisons (Tables 1, S5). We observed the strongest correlations between dispersal barrier and kinship for the Isthmus of Panama, Andes Mountains, Rio Negro, and Rio Tapajós.

We discarded the first five million generations of all BEDASSLE MCMC chains and used the remaining posterior to estimate the ratio of the effect size of barriers versus the effect size of geographic distance ($\alpha_E/\alpha_D$). Across all barriers, the mean and median ratios were 0.413 and the 95% credible set was 0.322 to 0.464. The interpretation of this ratio is that the effect on genetic differentiation of separation by a barrier is equivalent to the effect of roughly 2,000 to 2,900 km of geographic distance. Examining each barrier separately, we found variation across barriers in the relative effect sizes of the barrier and geographic distance (Table 1). The Andes Mountains, Rio Negro, Rio Tapajós, and Cerrado Belt had the highest ratios, supporting the particular importance of these barriers in structuring genetic variation.

Population assignment and admixture

Analysis of P(X|D) from the STRUCTURE runs suggested K=5 was the optimal value for number of populations (Table S6). The five clusters from the K=5 analysis contained the individuals from (Central America + Chocó), Guiana, (Napo + Inambari + Rondônia), (Tapajós + Xingu), and Atlantic Forest (Figs. 3, S1). The four populations from the K=4 analysis were similar, except the Guiana population was lumped with the (Napo + Inambari + Rondônia) population (Fig. S1).

The Structurama results were sensitive to the specification of the α prior. The (1,1) prior, with a small mean and narrow s.d. resulted in three populations; the (5,1) prior with an intermediate mean and s.d. resulted in five populations; the (10,1) prior with a large mean and s.d. resulted in four populations; and the prior based on an expected value of nine populations resulted in three populations (Fig. S1). The populations from all analyses included some combination of the same populations from the STRUCTURE analysis. The five populations from the Structurama analysis with an intermediate prior of (5,1) were identical to the five populations from the STRUCTURE analysis at K=5 (Figs. 3, S1). These five populations were selected for use in subsequent analyses.

fineSTRUCTURE revealed more population structure than did STRUCTURE and Structurama. For $c = 1/(n-1)$, eight populations were detected (Figs. 3, S1, S2). These resembled the populations from the STRUCTURE analysis with K=5 and the Structurama analysis with the (5,1) prior. fineSTRUCTURE, however, divided the (Central America + Chocó) population into two, with the break occurring west of the canal zone in Panama (an individual from Coclé just west of the canal is allied with the Chocó individuals), and identified a cluster within Central America comprising the two northwestern-most samples from foothill areas in Oaxaca and Chiapas, Mexico. In addition, fineSTRUCTURE separated seven of the eight individuals in the Napo region from those in the Inambari and Rondônia regions. The eighth sample from the Napo region allied with the Inambari and Rondônia samples. This sample was collected in the foothills of southern Ecuador not far from the Río Marañon, which is often considered the border between the Napo and Inambari regions. Varying the value of $c$ within a narrow range did not strongly influence cluster assignment in fineSTRUCTURE, and did so in an intuitive manner (e.g. by combining two weakly divergent clusters). We selected the eight populations from the fineSTRUCTURE analysis with $c = 1/(n-1)$ for use in subsequent analyses.

Both the set of populations inferred from fineSTRUCTURE (r = -0.6709, p = 0.0001) and STRUCTURE/Structurama (r = -0.7611, p = 0.0001) explained kinship between individuals significantly, even after controlling for isolation by distance in partial Mantel tests (Table 1). An examination of the admixture estimates from the STRUCTURE analysis with K=5 revealed relatively low admixture between populations (Fig. S3). A small amount of admixture was observed between Guiana and (Napo + Inambari + Rondônia) and between (Napo + Inambari + Rondônia) and (Tapajós + Xingu).

Population expansion and migration

LAMARC MCMC chains converged after 2-3 million generations, but were run to 20 million. In both the analyses of fineSTRUCTURE and STRUCTURE/Structurama populations, θ was smaller in the Atlantic Forest population than in all other populations except the Napo population in the fineSTRUCTURE analysis (Table 2). We detected significant population growth (confidence intervals not overlapping zero) in seven of the eight fineSTRUCTURE populations and all five of the STRUCTURE/Structurama populations (Table 2). Growth rates were higher in the (Tapajós + Xingu) and Atlantic Forest populations than in other populations, except for the Central American and Guianan populations in the analysis of fineSTRUCTURE populations.

We recovered significant non-zero migration rates (confidence intervals not overlapping zero) in six of the 14 pairwise estimates for the fineSTRUCTURE populations and three of the eight pairwise estimates for the STRUCTURE/Structurama populations (Table 3). Migration between Central American and Mexican populations in the analysis of fineSTRUCTURE populations was higher than between most other populations. Migration was also detected from the Chocó region to Central America (fineSTRUCTURE), from the (Napo + Inambari + Rondônia) population to the trans-Andean populations (STRUCTURE/Structurama), and from the (Tapajós + Xingu) population to the Atlantic Forest (both analyses). Within the Amazon

Basin, analysis of the STRUCTURE/Structurama populations detected migration in both directions across the Negro River, and analysis of the fineSTRUCTURE populations detected migration from the Napo to the Guianan and (Inambari + Rondônia) populations and from the (Inambari + Rondônia) population to the (Tapajós + Xingu) population.

Natural selection

We detected no loci putatively under diversifying selection using BayeScan with the STRUCTURE/Structurama populations and the false discovery rate (FDR) set to 0.05 (Fig. S3). We did, however, detect 20 loci that were putatively under purifying or balancing selection (FDR=0.05). In the analysis of the fineSTRUCTURE populations we detected 32 loci putatively under diversifying selection and 41 loci putatively under purifying or balancing selection (FDR=0.05). Of the 20 loci putatively under purifying/balancing selection in the analysis of STRUCTURE/Structurama populations, 17 were also outliers putatively under purifying/balancing selection in the analysis of fineSTRUCTURE populations.

Species tree

We recovered well-supported topologies from the SNAPP species tree analyses of both the STRUCTURE/Structurama population set and the fineSTRUCTURE population set (PP of all nodes = 1.0). Runs converged after two to three million generations, so we used a burnin of three million generations. We ran both runs for each set of populations an additional four million generations and used the combined sample of 4,000 trees to generate a Maximum Clade

Credibility tree and posterior probability values for each node (Fig. 4). Topologies were consistent between the analysis of the STRUCTURE/Structurama populations and the analysis of the fineSTRUCTURE populations. Both estimated an initial divergence between the Atlantic Forest population and all other populations, followed by a divergence across the Andes. Within the Amazon Basin, both analyses estimated an earlier divergence across the Tapajós River followed by a subsequent divergence across the Negro River. Divergences between the two Central American populations, the Central American and Chocó populations, and the Napo and (Inambari + Rondônia) populations from the fineSTRUCTURE analysis were very shallow.

The SNP species tree was similar overall to a prior mitochondrial gene tree based on Cytochrome b data from the same samples used in this study (Smith et al. in review, Fig. 4). It differed, however, in the placement of the Guianan population. In the SNP species trees, the Guianan population is sister to the (Napo + Inambari + Rondônia) clade with high support (PP = 1.0), and thus is nested within the clade containing the other Amazonian populations. In the mitochondrial gene tree, however, the Guianan population is sister, albeit with a very long intervening branch, to the Atlantic Forest population with high support (PP = 0.94).

Discussion

Prior studies of *Xenops minutus* based on mitochondrial sampling from many individuals (Burney 2009) or genomic sampling from a few individuals (Smith et al. 2014) revealed deep phylogeographic structure associated with major landscape features, such as the Andes mountains and Amazonian rivers. Our GBS data identified the same phylogeographic breaks. Moreover, our results indicate the historical demography of *X. minutus* has been dynamic, with population size changes, migration and admixture between populations, and possibly natural selection. These processes, along with local extinctions, re-colonization, and competition, all operating in a heterogeneous geographic landscape, produce the structuring and levels of genetic variation we can measure from modern population samples.

We recovered positive population growth estimates for nearly all populations in the LAMARC analysis. Growth was greater in the (Tapajós + Xingu) and Atlantic Forest populations in the southeastern portion of the distribution than in most other populations. Signatures of population growth have been observed in some other Neotropical forest species (Aleixo 2004; Cheviron et al. 2005; Solomon et al. 2008; D'Horta et al. 2011, but see Lessa et al. 2003). The significant migration rates and evidence of admixture confirm that connectivity between currently isolated populations has occurred over the history of *X. minutus*. We recovered significant non-zero estimates for 9 of 22 total migration parameters across two different analyses in LAMARC. Across the Andes Mountains and cerrado belt, we detected significant migration in only one direction - out of rather than into the Amazon Basin. The STRUCTURE analysis also suggested the presence of limited admixture in some populations. In addition, we directly identified an admixed individual: the individual from the Napo region that clustered with the Inambari SNP clade. Prior mitochondrial data from this individual (Burney 2009) reveals a haplotype that clusters closely with other Napo individuals, rather than individuals from the Inambari region (Fig. 4). This admixed individual therefore has a Napo mitochondrial haplotype, but an Inambari nuclear SNP genotype. There are few previous estimates of migration rate between populations of Neotropical forest organisms isolated by barriers, and these mostly suggest that gene flow is low or absent (Patton et al. 1994; Noonan and Gaucher 2005; Maldonado-Coelho et al. 2013).

Hybridization and introgression between species and divergent forms have been uncovered in a few Neotropical taxa (Brumfield et al. 2001; Lovette 2004; Dasmahapatra et al. 2010; Naka et al. 2012). We expect that increased genomic representation in datasets will reveal that migration, hybridization, and introgression are an important part of the diversification history of the Neotropics.

Although we detected a small proportion of loci under purifying or balancing selection, the detection and interpretation of loci under purifying or balancing selection (ie. lower divergence than expected) is challenging (Teacher et al. 2013) due to the diversity of processes that might underlie such a pattern. The detection of diversifying selection at a small proportion of loci in the BayeScan analysis of fineSTRUCTURE populations, but not in the analysis of STRUCTURE/Structurama populations, suggested that diversifying selection has occurred between the most recently diverged populations. We found, however, that none of the outliers putatively under diversifying selection showed large allele frequency differences between populations that were only separated in the fineSTRUCTURE population set. Null $F_{st}$ distributions may be overly narrow when some populations are recently diverged and have highly correlated allele frequencies, resulting in false positive outliers (Excoffier et al. 2009). Correlated allele frequencies between recently diverged populations in the fineSTRUCTURE analysis, rather than diversifying selection, are likely responsible for the positive outliers in that analysis.

Accurately mapping loci to an annotated genome assembly may permit further evaluation of putative outliers (Stapley et al. 2010), but is complicated in our study by the absence of a genome assembly for *X. minutus* or any close relative, as well as the short length (~64 bp) of the GBS loci. Because we lack an independent method of verifying outliers, our results are very preliminary with regards to the importance of selection in this system. In addition to the problems mentioned above, the total number of loci putatively under selection across both BayeScan analyses (76 loci, 2.2% of the total) is smaller than in many other studies (reviewed in Nosil et al. 2009), suggesting a relatively minor role for selection in the history of *X. minutus*.

Relating species history to landscape history is challenging

Although we recovered a detailed estimate of the history of *X. minutus*, relating this history to the landscape history of the Neotropics and to hypotheses of Neotropical diversification in general is challenging. Similar issues have been encountered in other studies, such that few general patterns have emerged that convincingly relate landscape history to diversification history within species (Antonelli et al. 2010; Brumfield 2012). The difficulty stems in part from the incomplete knowledge of Neotropical landscape history on spatial and temporal scales relevant for species evolution (Bush 1994; Bush and Flenley 2007) and from the shortage of unique testable predictions under different hypotheses of Neotropical diversification (Brumfield and Capparella 1996; Tuomisto and Ruokolainen 1997). Another challenge is that species distributions appear to be dynamic on much shorter timescales than those on which landscape evolution occurs, potentially erasing the signal for important events and resulting in pseudo-congruence (Haydon et al. 1994; Sanmartin et al. 2008; Brumfield 2012). Finally, different species are likely to have responded in different ways to the same history depending on their ecologies, such that few general patterns may exist (Aleixo 2006, Rull 2013).

We did find that major Neotropical landscape features, including the Andes, Amazonian rivers, and the cerrado belt isolating Amazonia from the Atlantic Forests, accounted for much of the genetic structure within *X. minutus*. The species tree topology for *X. minutus* contains similar area relationships to those found in other phylogenetic analyses (Weckstein and Fleischer 2005, Aleixo and Rossetti 2007). Divergence across barriers may be evidence of vicariance associated with barrier origin, dispersal across an existing barrier followed by differentiation (Mayr 1963), or the role of the barrier in structuring variation that arose elsewhere due to unknown historical processes (Brumfield 2012). The potential for pseudo-congruence between barriers and distributions combined with recent evidence that dispersal is more important than vicariance in the histories of some Neotropical groups (Fine et al. in press, Smith et al. in review) suggests that the null hypothesis of shared area relationships used in vicariance biogeography is inappropriate. In addition, existing hypotheses of Neotropical diversification include few explicit predictions about relationships between areas of endemism (Bates et al. 1998, Leite and Rogers 2013), and replicate simulations illustrate a remarkable amount of phylogenetic discordance even under identical vicariance scenarios (Endler 1983). Because of these issues, the divergence patterns in *X. minutus* tell us relatively little about the historical landscape or climatic events responsible for the modern genetic structuring in this species.

Dating the divergences between populations could allow determination of whether they were coincident with barrier formation, providing circumstantial support for particular vicariance hypotheses. Although dating the SNP divergences is problematic because we lack substitution rate estimates for GBS loci (see below), a previous dating analysis using mitochondrial DNA suggested that *X. minutus* populations diverged within the time span that the Andes Mountains and Amazonian Rivers are thought to have reached their modern conformations (Smith et al. in review). *Xenops minutus* populations across the Andes diverged 4.58 (s.d.=3.04-5.98) Mya and populations within the Amazon basin (aside from the Guianan population with a potential spurious placement in the mitochondrial tree, see below) began diverging 2.91 (s.d.=1.89-4.00)

Mya. Similar Pliocene divergence dates have been estimated for many other Neotropical taxa including fish (e.g., Lovejoy et al. 2010; Lundberg et al. 2010), plants (e.g., Pennington and Dick 2010), amphibians (e.g., Santos et al. 2009), birds (e.g., Weir 2006), and mammals (e.g., Costa 2003). These dates coincide roughly with the final uplift of the Andes and the coincident formation of the contemporary fluvial system of the Amazon in the last 10 My (Mora et al. 2010). However, the concordance of divergence dates with the vast time span associated with the origin of these dispersal barriers provides only rough, circumstantial support. The crucial details of how dispersal barriers interdigitate with other factors, such as population size flux, changes in forest distribution (Bush and Flenley 2007), changes in forest composition and niche availability (Jaramillo et al. 2010), changes in avian community composition (Ricardo Negri et al. 2010), and local extinctions and re-colonizations are not considered. This uncertainty suggests a nuanced understanding of how the Andes and Amazonian rivers influence speciation within lineages is not achievable using area relationships and divergence dates, and that our focus should be on other aspects of the speciation process.

The evidence we found for population expansions in *X. minutus* provides support for a prediction of the forest refugia hypothesis that humid lowland forests were once more restricted due to the expansion of savanna (Haffer 1969). Some palynological analyses also support the idea that lowland Neotropical humid forest was once more restricted (Absy et al. 1991; Burnham and Graham 1999). Recent isotopic evidence suggests that precipitation was lower in the eastern Amazon, but not the western Amazon, during the last glaciation (Cheng et al. 2013), consistent with our observation of greater population growth in that area. Unfortunately, knowledge of the recent history of forest cover in the Amazon is limited and contentious (Behling et al. 2010). The marine incursion hypotheses might also predict population growth following the recession of water levels, although growth is expected to be greatest in the western Amazon Basin (Aleixo 2004), contrary to the pattern we observed. Other events such as disease (e.g., Daszak et al. 2003), changes in abiotic climate conditions (e.g., Sillett et al. 2000), or changes in competitive interactions (e.g., Koenig 2003), predation (e.g., Wittmer et al. 2005), or resource availability (e.g., O'Donoghue et al. 1997) might also have driven population size changes. Although the population expansion we observed in *X. minutus* may be attributable to recent increases in forest habitat in the lowland Neotropics, we cannot exclude other equally likely causes.

Migration and admixture between populations supports the idea that populations have experienced periodic connections in the past. Habitat connectivity, however, might have occurred under any of various hypotheses of Neotropical diversification and does not aid in discriminating among them. Future improvements in our understanding of past habitat distributions combined with improved methods of inferring and dating admixture events may allow us to correlate episodes of migration and gene flow with individual events of habitat connectivity (Gillespie and Roderick 2014).

Based on the challenges associated with connecting the species history of *X. minutus* to landscape history, we suggest the common practice of relating single species histories to landscape events is unproductive. As an alternative, researchers should focus on evaluating the importance of different historical processes (including divergence, but also population size changes, migration, and selection) across many datasets, and then use comparative methods to determine the importance of each process along taxonomic, temporal, and spatial axes. If this can be accomplished and if better resolved landscape histories can be estimated, it may eventually be possible to evaluate the importance of each hypothesis of Neotropical diversification across whole assemblages, timescales, and regions.

Limitations and prospects for GBS data in phylogeography

Genotyping-by-sequencing data allowed us to conduct a variety of population genetic, phylogeographic, and phylogenetic analyses. We did, however, encounter some potential shortcomings of GBS data for addressing phylogeographic questions in our non-model system. The large amount of missing data observed in our dataset prior to filtering suggests the need for further optimization of coverage relative to the number of targeted loci, but better coverage could be achieved by using different enzymes or multiple enzymes (Peterson et al. 2012). The locations to which we were able to map loci may be inaccurate, both because of the potential for spurious alignment due to the short length of the GBS reads, and because of the evolutionary distance between *X. minutus* and *T. guttata*. This issue may be reduced in the future if longer read lengths can be obtained, or if a genome from a species more closely related to the study species becomes available. Perhaps the greatest limitation of GBS is that no standard evolutionary rate exists for the targeted loci for the purpose of dating divergences or converting demographic parameters. As a result, we were largely limited to making relative comparisons of raw parameter estimates in this study. Furthermore, the processing pipeline for GBS and other RAD-Seq data complicates the future development of standard rates that could be used across groups of organisms. Because identity thresholds are applied to each dataset for assembly, datasets may be truncated to different degrees and rates are not directly comparable. More informed assembly protocols or methods for correcting rates based on the level of truncation in a dataset may alleviate these issues in the future.

Despite some limitations, genomic data from GBS have provided a more complete picture of the history of *X. minutus* than would be possible with a few markers. The history inferred from genomic SNPs is likely to better reflect the true history of *X. minutus* populations than a single-locus dataset (Edwards and Beerli 2000). In addition, genomic data have allowed us to investigate processes that are difficult to evaluate with a single marker, such as migration and selection. More efficient laboratory methods and new analytical tools will surely increase the utility of genomic datasets as they come into more widespread use.

Since divergence histories based on mitochondrial data have been the primary source of information for studies of Neotropical phylogeography (Haffer 1997; Antonelli et al. 2010; Leite and Rogers 2013), the discordance between the mitochondrial gene tree and genome-wide SNP species trees in this study is alarming. The source of this discrepancy is unclear, but it is possible that deep coalescence of the mitochondrial haplotypes from the Guianan and Atlantic Forest populations resulted in a mitochondrial genealogy that does not represent the species history. This result is not surprising, given the number of prior studies reporting discordance between mitochondrial and large nuclear datasets (Funk and Omland 2003; Chan and Levin 2005). The observed discordance argues for careful interpretation of mitochondrial data and the importance of shifting to genome-wide datasets for phylogeographic research.

Acknowledgements

We gratefully acknowledge the following people and institutions for providing samples: Paul Sweet, Thomas J. Trombone, and Joel Cracraft (AMNH); Nathan H. Rice (ANSP); Kimberly

Bostwick and Charles Dardia (CUMV); David E. Willard, Shannon J. Hackett, and Jason D. Weckstein (FMNH); Mark B. Robbins (KUMNH); Donna L. Dittmann (LSUMZ); Alexandre Aleixo (MPEG), Adolfo G. Navarro-Sigüenza (MZFC-UNAM); Luis Fabio Silveira (MZUSP); Gary R. Graves and James Dean (USNM); John Klicka and Sharon Birks (UWBM). Scott A. Taylor and Thomas A. White provided assistance with GBS data processing. Gideon Bradburd provided help with running BEDASSLE and Brian Tilston Smith assisted with various analyses. J. V. Remsen, Jr. and the LSU bird lunch group provided helpful comments. Funding was provided by an NSF Doctoral Dissertation Improvement Grant to M.G.H. (DEB-1210556).

References

Absy M, Cleef A, Fournier M, *et al*. (1991) Mise en évidence de quatre phases d'ouverture de la forêt dense dans le sud-est de l'Amazonie au cours des 60 000 dernières années. Première comparaison avec d'autres régions tropicales. *Comptes rendus de l'Académie des sciences Paris*, **312**, 673-678.

Aleixo A (2004) Historical diversification of a terra-firme forest bird superspecies: a phylogeographic perspective on the role of different hypotheses of Amazonian diversification. *Evolution,* **58**, 1303-1317.

Aleixo A (2006) Historical diversification of floodplain forest specialist species in the Amazon: a case study with two species of the avian genus Xiphorhynchus (Aves: Dendrocolaptidae). *Biological Journal of the Linnean Society*, **89**, 383-395.

Aleixo A, Rossetti DF (2007) Avian gene trees, landscape evolution, and geology: towards a modern synthesis of Amazonian historical biogeography? *Journal of Ornithology,* **148**, 443-453.

Altschul SF, Gish W, Miller W, Myers EW, Lipman DJ (1990) Basic local alignment search tool. *Journal of Molecular Biology*, **215**, 403-410.

Antonelli A, Quijada-Mascareñas A, Crawford AJ, *et al.* (2010) Molecular Studies and Phylogeography of Amazonian Tetrapods and their Relation to Geological and Climatic Models. In: *Amazonia, Landscape and Species Evolution: A Look into the Past* (eds Hoorn C, Wesselingh FP). pp. 386-404. Wiley-Blackwell, London, UK.

Baird NA, Etter PD, Atwood TS, *et al.* (2008) Rapid SNP discovery and genetic mapping using sequenced RAD markers. *PLoS One,* **3**, e3376.

Bates JA, Hackett SJ, and J Cracraft (1998) Area-relationships in the Neotropical lowlands: An hypothesis based on raw distributions of passerine birds. *Journal of Biogeography*, **25**, 783-793.

Behling H, Bush M, Hooghiemstra H (2010) Biotic Development of Quaternary Amazonia: A Palynological Perspective. In: *Amazonia, Landscape and Species Evolution: A Look into the Past* (eds Hoorn C, Wesselingh FP). pp. 335-345. Wiley-Blackwell, London, UK.

Bradburd GS, Ralph PL, Coop GM (2013) Disentangling the effects of geographic and ecological isolation on genetic differentiation. *Evolution,* **67**, 3258-3273.

Bradbury PJ, Zhang Z, Kroon DE, *et al.* (2007) TASSEL: Software for association mapping of complex traits in diverse samples. *Bioinformatics*, **23**, 2633-2635.

Brawn JD, Collins T, Medina M, Bermingham E (1996) Associations between physical isolation and geographical variation within three species of Neotropical birds. *Molecular Ecology,* **5**, 33-46.

Brumfield RT (2012) Inferring the origins of lowland Neotropical birds. *Auk*, **129**, 367-376.

Brumfield RT, Capparella AP (1996) Historical diversification of birds in northwestern South America: a molecular perspective on the role of vicariant events. *Evolution,* **50**, 1607-1624.

Brumfield RT, Jernigan RW, McDonald DB, Braun MJ (2001) Evolutionary implications of divergent clines in an avian (*Manacus*: Aves) hybrid zone. *Evolution*, **55**, 2070-2087.

Bryant D, Bouckaert R, Felsenstein J, Rosenberg NA, RoyChoudhury A (2012) Inferring species trees directly from biallelic genetic markers: bypassing gene trees in a full coalescent analysis. *Molecular Biology and Evolution*, **29**, 1917-1932.

Burney CW (2009) Comparative Phylogeography of Neotropical Birds. PhD Dissertation. Louisiana State University, Baton Rouge, LA, 133 pp.

Burnham RJ, Graham A (1999) The history of neotropical vegetation: New developments and status. *Annals of the Missouri Botanical Garden*, **86**, 546-589.

Bush MB (1994) Amazonian speciation: A necessarily complex model. *Journal of Biogeography,* **21**, 5-17.

Bush MB, Flenley J (2007) Tropical Rainforest Responses to Climatic Change. Springer, New York, USA.

Cabanne GS, Santos FR, Miyaki CY (2007) Phylogeography of Xiphorhynchus fuscus (Passeriformes, Dendrocolaptidae): vicariance and recent demographic expansion in southern Atlantic forest. *Biological Journal of the Linnean Society*, **91**, 73-84.

Capparella AP (1987) Effects of riverine barriers on genetic differentiation of Amazonian forest undergrowth birds. PhD Dissertation. Louisiana State University, Baton Rouge, LA, 137 pp.

Carlson CS , Thomas DJ, Eberle MA, *et al.* (2005) Genomic regions exhibiting positive selection identified from dense genotype data. *Genome Research*, **15**, 1553-1565.

Chan K, Levin SA (2005) Leaky prezygotic isolation and porous genomes: rapid introgression of maternally inherited DNA. *Evolution*, **59**, 720-729.

Chapman FM (1917) The Distribution of Bird-life in Colombia: A Contribution to a Biological Survey of South America. *Bulletin of the American Museum of Natural History*, **36**, 1-729.

Chapman FM (1926) The distribution of bird-life in Ecuador. *Bulletin of the American Museum of Natural History*, **55**, 1-784.

Cheng H, Sinha A, Cruz FW, *et al*. (2013) Climate change patterns in Amazonia and biodiversity. *Nature Communications*, **4**, 1411.

Cheviron Z, Hackett SJ, Capparella AP (2005) Complex evolutionary history of a Neotropical lowland forest bird (*Lepidothrix coronata*) and its implications for historical hypotheses of the origin of Neotropical avian diversity. *Molecular Phylogenetics and Evolution*, **36**, 338-357.

Costa LP (2003) The historical bridge between the Amazon and the Atlantic Forest of Brazil: a study of molecular phylogeography with small mammals. *Journal of Biogeography*, **30**, 71-86.

Cracraft J (1985) Historical biogeography and patterns of differentiation within the South American avifauna: Areas of endemism. Ornithological Monographs, **36**, 49-84.

Cracraft J, Prum RO (1988) Patterns and processes of diversification: speciation and historical congruence in some Neotropical birds. *Evolution*, **42**, 603-620.

D'Horta FM, Cabanne GS, Meyer D, Miyaki CY (2011) The genetic effects of Late Quaternary climatic changes over a tropical latitudinal gradient: Diversification of an Atlantic Forest passerine. *Molecular Ecology*, **20**, 1923-1935.

Dasmahapatra KK, Lamas G, Simpson F, Mallet J (2010) The anatomy of a "suture zone" in Amazonian butterflies: A coalescent-based test for vicariant geographic divergence and speciation. *Molecular Ecology*, **19**, 4283-4301.

Daszak P, Cunningham AA, Hyatt, AD (2003) Infectious disease and amphibian population declines. *Diversity and Distributions*, **9**, 141-150.

Dickinson EC (2003) The Howard and Moore Complete Checklist of the Birds of the World. Christopher Helm, London, UK.

Edwards S, Beerli P (2000) Perspective: Gene divergence, population divergence, and the variance in coalescence time in phylogeographic studies. *Evolution,* **54**, 1839-1854.

Elshire RJ, Glaubitz JC, Sun Q, *et al.* (2011) A robust, simple genotyping-by-sequencing (GBS) approach for high diversity species. *PLoS One*, **6**, e19379.

Endler, JA (1983) Testing Causal Hypotheses in the Study of Geographical Variation. In: *Numerical Taxonomy* (ed Felsenstein J). pp. 424-443. Springer-Verlag, Berlin, Germany.

Erwin TL (1979) Thoughts on the Evolutionary History of Ground Beetles: Hypotheses Generated from Comparative Faunal Analyses of Lowland Forest Sites in Temperate and Tropical Regions. In: *Carabid Beetles, their Evolution, Natural History, and Classification* (eds Erwin TL, Ball GE, Whitehead DR, Halpern AL). pp. 539-592. DW Junk, The Hague, Netherlands.

Excoffier L, Hofer T, Foll M (2009) Detecting loci under selection in a hierarchically structured population. *Heredity,* **103**, 285-298.

Falush D, Stephens M, Pritchard JK (2003) Inference of population structure using multilocus genotype data: linked loci and correlated allele frequencies. *Genetics,* **164**, 1567-1587.

Fine PVA, Zapata F, Daly DC (In press) Investigating processes of Neotropical rain forest tree diversification by examining the evolution and historical biogeography of the Protieae (Burseraceae). *Evolution*.

Foll M, Gaggiotti O (2008) A genome-scan method to identify selected loci appropriate for both dominant and codominant markers: a Bayesian perspective. *Genetics*, **180**, 977-993.

Frantz LAF, Schraiber JG, Madsen O, *et al.* (2013) Genome sequencing reveals fine scale diversification and reticulation during speciation in *Sus. Genome Biology*, **14**, R107.

Funk DJ, Omland KE (2003) Species-level paraphyly and polyphyly: Frequency, causes, and consequences, with insights from animal mitochondrial DNA. *Annual Review of Ecology, Evolution, and Systematics*, **34**, 397-423.

Gillespie, RG, Roderick GK (2014) Evolution: Geology and climate drive diversification. *Nature*, **509**, 297-298.

Guillot G, Rousset F (2013) Dismantling the Mantel tests. *Methods in Ecology and Evolution*, **4**, 336-344.

Goslee SC, Urban DL. 2007. The ecodist package for dissimilarity-based analysis of ecological data. *Journal of Statistical Software*, **22**, 1-19.

Hackett, SJ, Kimball RT, Reddy S, *et al.* (2008) A phylogenomic study of birds reveals their evolutionary history. *Science,* **320**, 1763-1768.

Haffer J (1969) Speciation in Amazonian forest birds. *Science,* **165**, 131-137.

Haffer J (1997) Alternative models of vertebrate speciation in Amazonia: an overview. *Biodiversity Conservation*, **6**, 451-476.

Hall JP, Harvey DJ (2002) The phylogeography of Amazonia revisited: new evidence from riodinid butterflies. *Evolution,* **56**, 1489-1497.

Hardy OJ, Vekemans X (2002) SPAGeDi: A versatile computer program to analyse spatial genetic structure at the individual or population levels. *Molecular Ecology Notes,* **2**, 618-620.

Haydon DT, Radtkey RR, Pianka ER (1994) Experimental Biogeography: Interactions Between Stochastic, Historical, and Ecological Processes in a Model Archipelago. In: *Species Diversity in Ecological Communities: Historical and Geographical* Perspectives (eds Ricklefs RE, Schluter D). pp. 117-130. University of Chicago Press, Chicago, USA.

Hohenlohe PA, Bassham S, Etter PD, Stiffler N, Johnson EA, Cresko WA (2010) Population genomics of parallel adaptation in threespine stickleback using sequenced RAD tags. *PLoS Genetics,* **6**, e1000862.

Huelsenbeck JP, Andolfatto P, Huelsenbeck, ET (2011) Structurama: Bayesian inference of population structure. *Evolutionary Bioinformatics Online*, **7**, 55-59.

Jaramillo C, Hoorn C, Silva S, *et al.* (2010) The Origin of the Modern Amazon Rainforest: Implications of the Palynological and Palaeobotanical Record. In: *Amazonia, Landscape and Species Evolution: A Look into the Past* (eds Hoorn C, Wesselingh FP). pp. 317-334. Wiley-Blackwell, London, UK.

Koenig WD (2003) European Starlings and their effect on native cavity-nesting birds. *Conservation Biology*, **17**, 1134-1140.

Kuhner MK (2006) LAMARC 2.0: Maximum likelihood and Bayesian estimation of population parameters. *Bioinformatics,* **22**, 768-770.

Kuhner MK (2009) Coalescent genealogy samplers: windows into population history. *Trends in Ecology and Evolution*, **24**, 86-93.

Lawson DJ, Hellenthal G, Myers S, Falush D (2012) Inference of population structure using dense haplotype data. *PLoS Genetics*, **8**, e1002453.

Legendre P, Fortin MJ (2010) Comparison of the Mantel test and alternative approaches for detecting complex multivariate relationships in the spatial analysis of genetic data. *Molecular Ecology Resources,* **10**, 831-844.

Leite RN, Rogers DS (2013) Revisiting Amazonian phylogeography: Insights into diversification hypotheses and novel perspectives. *Organisms, Diversity, and Evolution*, **13**, 639-664.

Lessa EP, Cook JA, Patton JL (2003) Genetic footprints of demographic expansion in North America, but not Amazonia, during the Late Quaternary. *Proceedings of the National Academy of Sciences*, **100**, 10331-10334.

Li H, Durbin R (2011) Inference of human population history from individual whole-genome sequences. *Nature*, **475**, 493-496.

Loiselle BA, Sork VL, Nason J, Graham C (1995) Spatial genetic structure of a tropical understory shrub, *Psychotria officinalis* (Rubiaceae). *American Journal of Botany*, **82**, 1420-1425.

Lougheed S, Gascon C, Jones D, Bogart J, Boag P (1999) Ridges and rivers: a test of competing hypotheses of Amazonian diversification using a dart-poison frog (*Epipedobates femoralis*). *Proceedings of the Royal Society B: Biological Sciences*, **266**, 1829-1835.

Lovejoy NR, Willis SC, Albert, JS (2010) Molecular Signatures of Neogene Biogeographical Events in the Amazon Fish Fauna. In: *Amazonia, Landscape and Species Evolution: A Look into the Past* (eds Hoorn C, Wesselingh FP). pp. 405-417. Wiley-Blackwell, London, UK.

Lovette IJ (2004) Molecular phylogeny and plumage signal evolution in a trans Andean and circum Amazonian avian species complex. *Molecular Phylogenetics and Evolution*, **32**, 512-523.

Lundberg JG, Sabaj Pérez MH, Dahdul WM, Aguilera OA (2010) The Amazonian Neogene Fish Fauna. In: *Amazonia, Landscape and Species Evolution: A Look into the Past* (eds Hoorn C, Wesselingh FP). pp. 281-301. Wiley-Blackwell, London, UK.

Lynch M (2009) Estimation of allele frequencies from high-coverage genome-sequencing projects. *Genetics*, **182**, 295-301.

Maldonado-Coelho M, Blake J, Silveira L, Batalha-Filho H, Ricklefs R (2013) Rivers, refuges and population divergence of fire-eye antbirds (*Pyriglena*) in the Amazon Basin. *Journal of Evolutionary Biology,* **26**, 1090-1107.

Mallet J (1993) Speciation, Raciation, and Color Pattern Evolution in *Heliconius* Butterflies: Evidence from Hybrid Zones. In: *Hybrid Zones and the Evolutionary Process* (eds Harrison RG). pp. 226-260. Oxford University Press, Oxford, UK.

Mayr, E (1963) Animal Species and Evolution. Belknap Press, Cambridge, USA.

Meirmans PG (2012) The trouble with isolation by distance. *Molecular Ecology,* **21**, 2839-2846.

Mora A, Baby P, Roddaz M, *et al.* (2010) Tectonic History of the Andes and Sub-Andean Zones: Implications for the Development of the Amazon Drainage Basin. In: *Amazonia,*

*Landscape and Species Evolution: A Look into the Past* (eds Hoorn C, Wesselingh FP). pp. 38-60. Wiley-Blackwell, London, UK.

Moritz C, Patton J, Schneider C, Smith T (2000) Diversification of rainforest faunas: an integrated molecular approach. *Annual Review of Ecology and Systematics*, **31**, 533-563.

Nadeau NJ, Martin SH, Kozak KM, *et al.* (2013) Genome-wide patterns of divergence and gene flow across a butterfly radiation. *Molecular Ecology,* **22**, 814-826.

Naka LN, Bechtoldt CL, Henriques LMP, Brumfield RT (2012) The role of physical barriers in the location of avian suture zones in the Guiana Shield, northern Amazonia. *American Naturalist*, **179**, E115-E132.

Noonan BP, Gaucher P (2005) Phylogeography and demography of Guianan harlequin toads (*Atelopus*): diversification within a refuge. *Molecular Ecology*, **14**, 3017-3031.

Nores M (1999) An alternative hypothesis for the origin of Amazonian bird diversity. *Journal of Biogeography,* **26**, 475-485.

Nosil P, Funk DJ, Ortiz-Barrientos D (2009) Divergent selection and heterogeneous genomic divergence. *Molecular Ecology*, **18**, 375-402.

O'Donoghue M, Boutin S, Krebs CJ, Hofer EJ (1997) Numerical responses of coyotes and lynx to the snowshoe hare cycle. *Oikos*, **80**, 150-162.

Patton JL, Da Silva MNF, Malcolm JR (1994) Gene genealogy and differentiation among arboreal spiny rats (Rodentia: Echimyidae) of the Amazon basin: a test of the riverine barrier hypothesis. *Evolution,* **48**, 1314-1323.

Patton JL, Da Silva MNF, Malcolm JR (2000) Mammals of the Rio Juruá and the evolutionary and ecological diversification of Amazonia. *Bulletin of the American Museum of Natural History*, **244**, 1-306.

Pearson DL (1977) A pantropical comparison of bird community structure on six lowland forest sites. *Condor*, **79**, 232-244.

Pennington RT, Dick CW (2010) Diversification of the Amazonian Flora and its Relation to Key Geological and Environmental Events: A Molecular Perspective. In: *Amazonia, Landscape and Species Evolution: A Look into the Past* (eds Hoorn C, Wesselingh FP). pp. 373-385. Wiley-Blackwell, London, UK.

Peterson BK, Weber JN, Kay EH, Fisher HS, Hoekstra HE (2012) Double digest RADseq: An inexpensive method for de novo SNP discovery and sequencing in model and non-model species. *PLoS One*, **7**, e37135.

Pool JE, Hellmann I, Jensen JD, Nielsen R (2010) Population genetic inference from genomic sequence variation. *Genome Research*, **20**, 291-300.

Pritchard JK, Stephens M, Donnelly P (2000) Inference of population structure using multilocus genotype data. *Genetics*, **155**, 945-959.

Quijada-Mascareñas JA, Ferguson JE, Pook CE, *et al*. (2007) Phylogeographic patterns of trans-Amazonian vicariants and Amazonian biogeography: the Neotropical rattlesnake (*Crotalus durissus* complex) as an example. *Journal of Biogeography,* **34**, 1296-1312.

Rambaut A, Drummond A (2007) Tracer v1.4. Available from http://beast.bio.ed.ac.uk/Tracer.

Rambaut A, Drummond A (2008) TreeAnnotator v1.4.8. Available from http://beast.bio.ed.ac.uk/TreeAnnotator.

Remsen JV (2003) Family Furnariidae (Ovenbirds). In: *Handbook of the Birds of the World* (eds del Hoyo J, *et al.*). pp. 162-357. Lynx Edicions, Barcelona, Spain.

Ribas CC, Aleixo A, Nogueira AC, Miyaki CY,  Cracraft J (2012) A palaeobiogeographic model for biotic diversification within Amazonia over the past three million years. *Proceedings of the Royal Society B: Biological Sciences*, **279**, 681-689.

Ricardo Negri F, Bocquentin-Villanueva J, Ferigolo J, Antoine PO (2010) A Review of Tertiary Mammal Faunas and Birds from Western Amazonia. In: *Amazonia, Landscape and Species Evolution: A Look into the Past* (eds Hoorn C, Wesselingh FP). pp. 243-258. Wiley-Blackwell, London, UK.

Rull V (2013) Some problems in the study of the origin of Neotropical biodiversity using palaeoecological and molecular phylogenetic evidence. *Systematics and Biodiversity,* **11**, 415-423.

Sanmartín I, Van Der Mark P, Ronquist F (2008) Inferring dispersal: A Bayesian approach to phylogeny-based island biogeography, with special reference to the Canary Islands. *Journal of Biogeography*, **35**, 428-449.

Santos JC, Coloma LA, Summers K, *et al.* (2009) Amazonian amphibian diversity is primarily derived from late Miocene Andean lineages. *PLoS Biology*, **7**, e1000056.

Sick H (1967) Rios e Enchentes na Amazônia como Obstáculo para a Avifauna. In: *Atas do Simpósio sobre a Biota Amazônica* (ed H. Lent). pp. 495-520. Conselho de Pesquisas, Rio de Janeiro, Brazil.

Sillett TS, Holmes RT, Sherry TW (2000) Impacts of a global climate cycle on population dynamics of a migratory songbird. *Science,* **288**, 2040-2042.

Smith BT, Harvey MG, Faircloth BC, Glenn TC, Brumfield RT (2014) Target capture and massively parallel sequencing of ultraconserved elements for comparative studies at shallow evolutionary time scales. *Systematic Biology*, **63**, 83-95.

Smith BT, McCormack JE, Cuervo AM, *et al*. (In review) The drivers of tropical speciation.

Solomon SE, Bacci M Jr, Martins M Jr, Vinha GG, Mueller UG (2008) Paleodistributions and comparative molecular phylogeography of leafcutter ants (*Atta* spp.) provide new insight into the origins of Amazonian diversity. *PLoS One,* **3**, e2738.

Stapley J, Reger J, Feulner PG, *et al.* (2010) Adaptation genomics: The next generation. *Trends in Ecology and Evolution*, **25**, 705-712.

Takahata N, Satta Y, Klein J (1995) Divergence time and population size in the lineage leading to modern humans. *Theoretical Population Biology*, **48**, 198-221.

Teacher AG, André C, Jonsson PR, Merilä J (2013) Oceanographic connectivity and environmental correlates of genetic structuring in Atlantic herring in the Baltic sea. *Evolutionary Applications*, **6**, 549-567.

Tuomisto H, Ruokolainen K (1997) The role of ecological knowledge in explaining biogeography and biodiversity in Amazonia. *Biodiversity Conservation*, **6**, 347-357.

Turner T, McPhee M, Campbell P, Winemiller K (2004) Phylogeography and intraspecific genetic variation of prochilodontid fishes endemic to rivers of northern South America. *Journal of Fish Biology*, **64**, 186-201.

Warren WC, Clayton DF, Ellegren H *et al*. (2010) The genome of a songbird. *Nature*, **464**, 757-762.

Weckstein JD, Fleischer R (2005) Molecular phylogenetics of the Ramphastos toucans: implications for the evolution of morphology, vocalizations, and coloration. *Auk*, **122**, 1191-1209.

Weir JT (2006) Divergent timing and patterns of species accumulation in lowland and highland neotropical birds. *Evolution,* **60**, 842-855.

Wesselingh F, Salo J (2006) A Miocene perspective on the evolution of the Amazonian biota. *Scripta Geologica,* **133**, 439-458.

White TA, Perkins SE, Heckel G, Searle JB (2013) Adaptive evolution during an ongoing range expansion: The invasive bank vole (*Myodes glareolus*) in Ireland. *Molecular Ecology*, **22**, 2971-2985.

Wittmer HU, Sinclair AR, McLellan BN (2005) The role of predation in the decline and extirpation of woodland caribou. *Oecologia*, **144**, 257-267.

Data Accessibility

Genotype data: Dryad (pending submission)

Custom scripts: Github (https://github.com/mgharvey)

Author Contributions

M.G.H. and R.T.B. designed the study and wrote the paper. M.G.H. conducted laboratory work, wrote custom scripts, and analyzed the data.

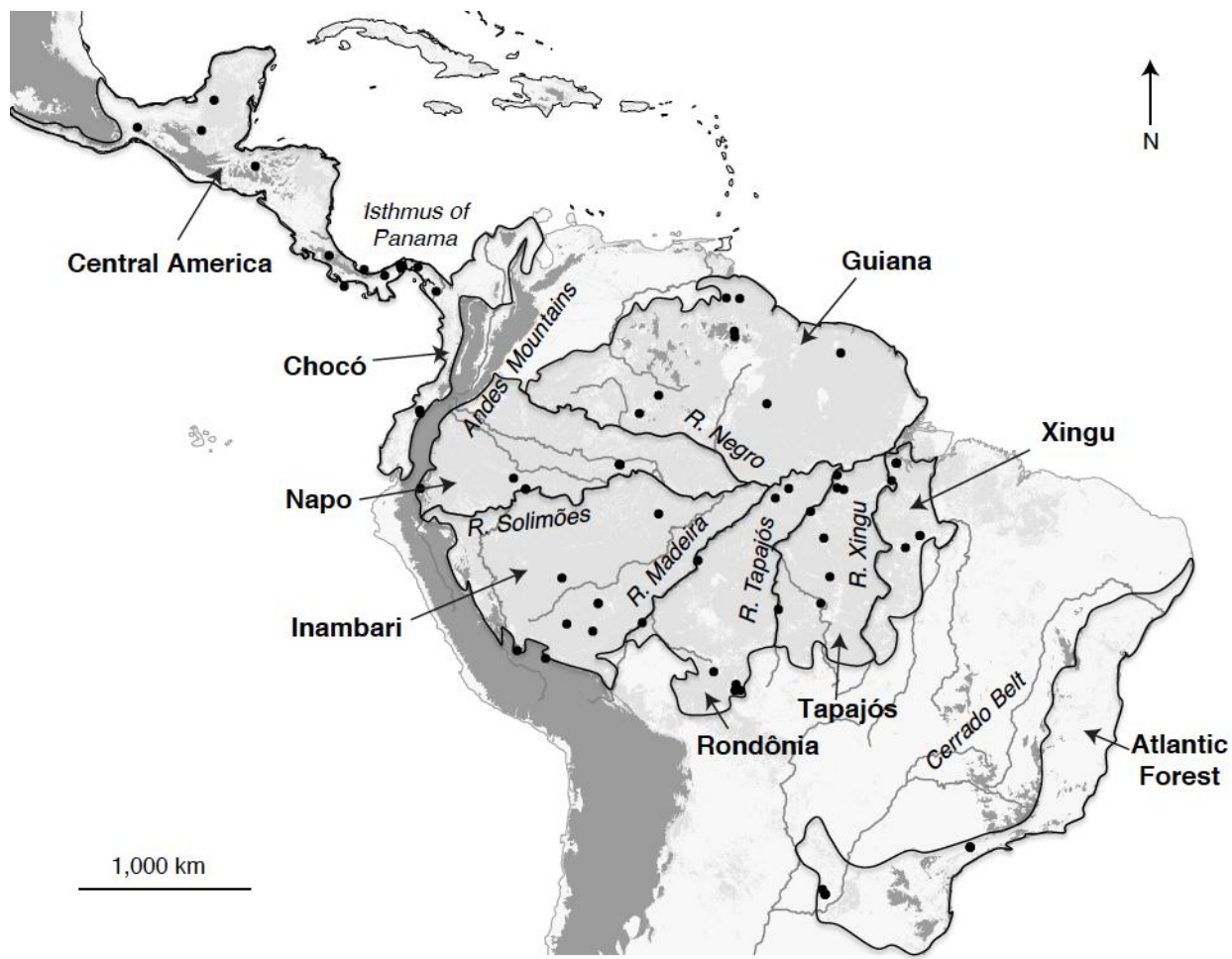

**Fig. 1.** Map showing sampling locations (circles), biogeographic areas (bold type) and dispersal barriers (italics) examined in this study.

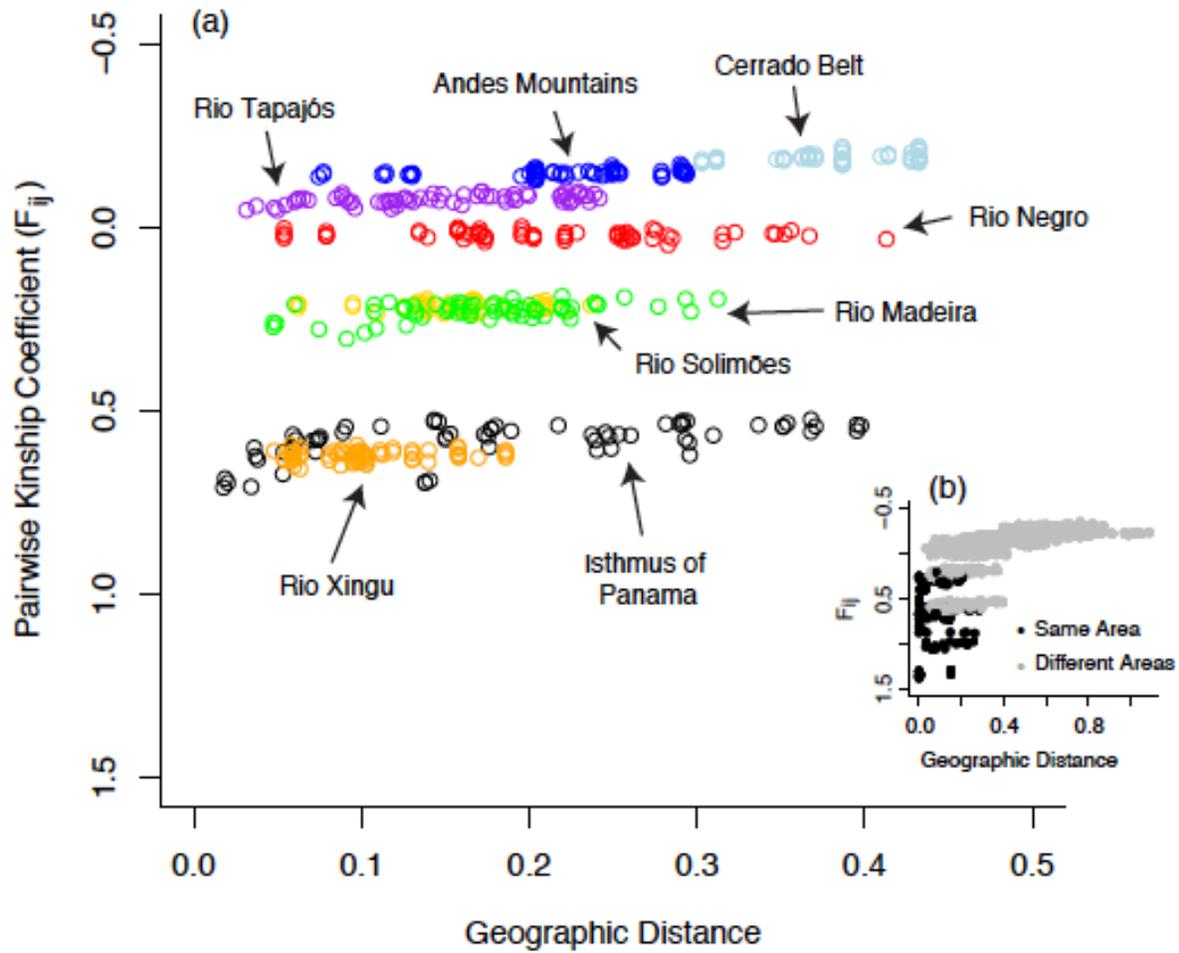

**Fig. 2.** Plots of pairwise kinship versus relative geographic distance (a) between individuals separated by a single putative barrier and (b) between all individuals including those within the same area (black points) or separated by one or multiple barriers (gray points). The y-axes are inverted so that points representing greater divergence appear toward the tops of the plots.

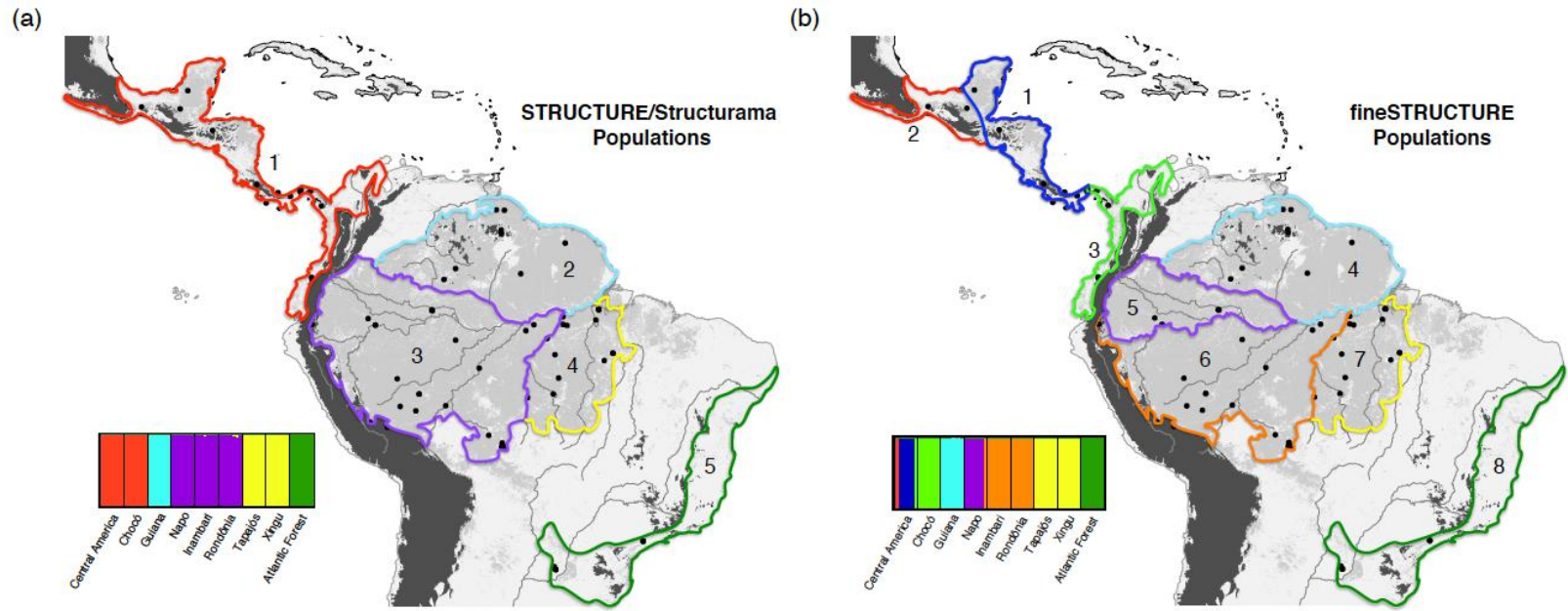

**Fig. 3.** Maps of the distributions of populations from (a) the STRUCTURE/Structurama analysis and (b) the fineSTRUCTURE analysis. Populations are numbered and numbers are consistent with subsequent tables and figures. The adjacent structure plots show population membership for all individuals from (a) the STRUCTURE analysis with K=5 and (b) the fineSTRUCTURE analysis. Admixed individuals are shown in the structure plot for the STRUCTURE analysis, but fineSTRUCTURE does not estimate admixture.

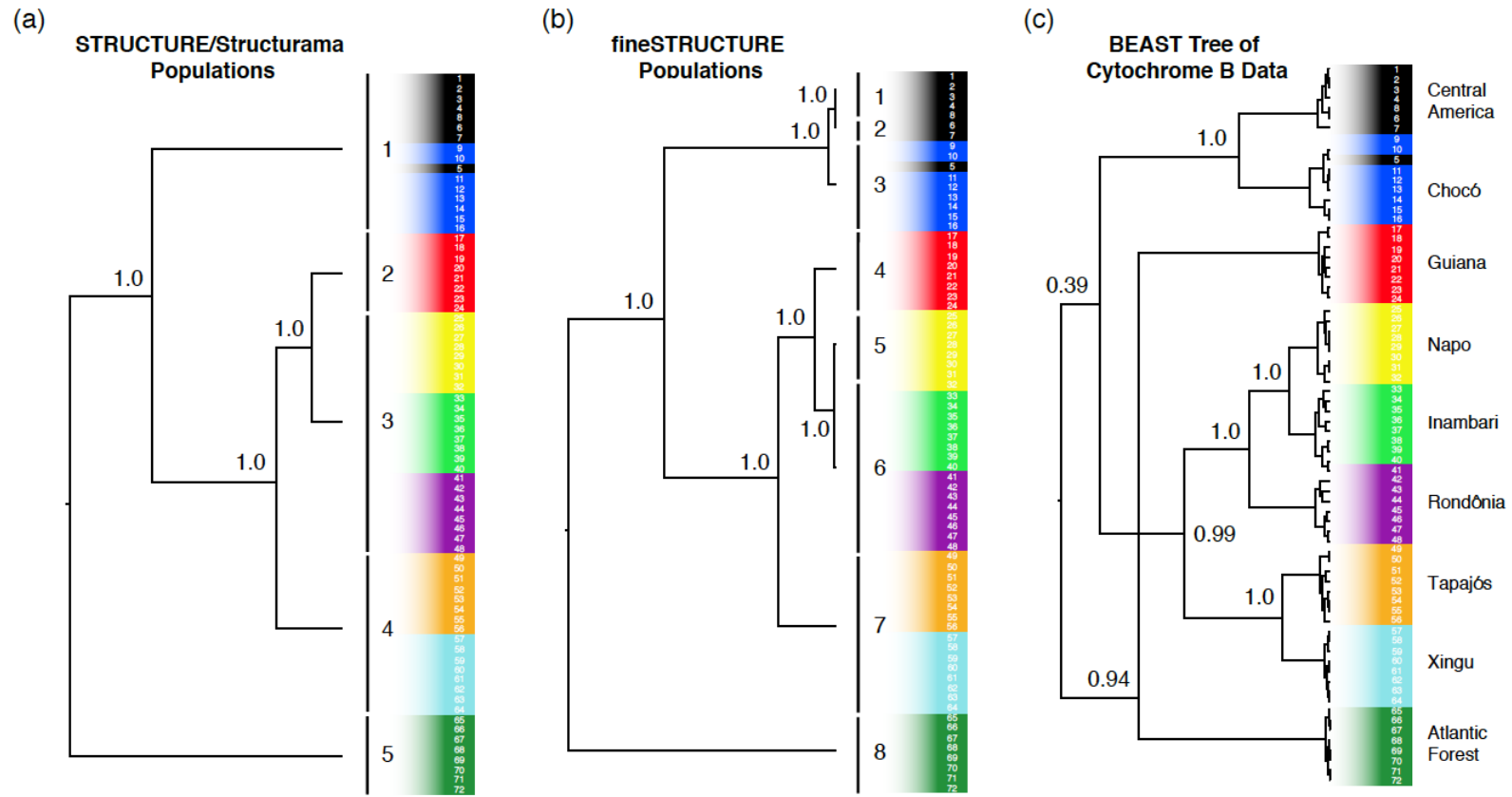

**Fig. 4** SNAPP species trees of (a) STRUCTURE/Structurama populations and (b) fineSTRUCTURE populations based on the SNP data and a (c) BEAST gene tree of sequence data from the mitochondrial gene Cytochrome B showing discordance with respect to the species trees.

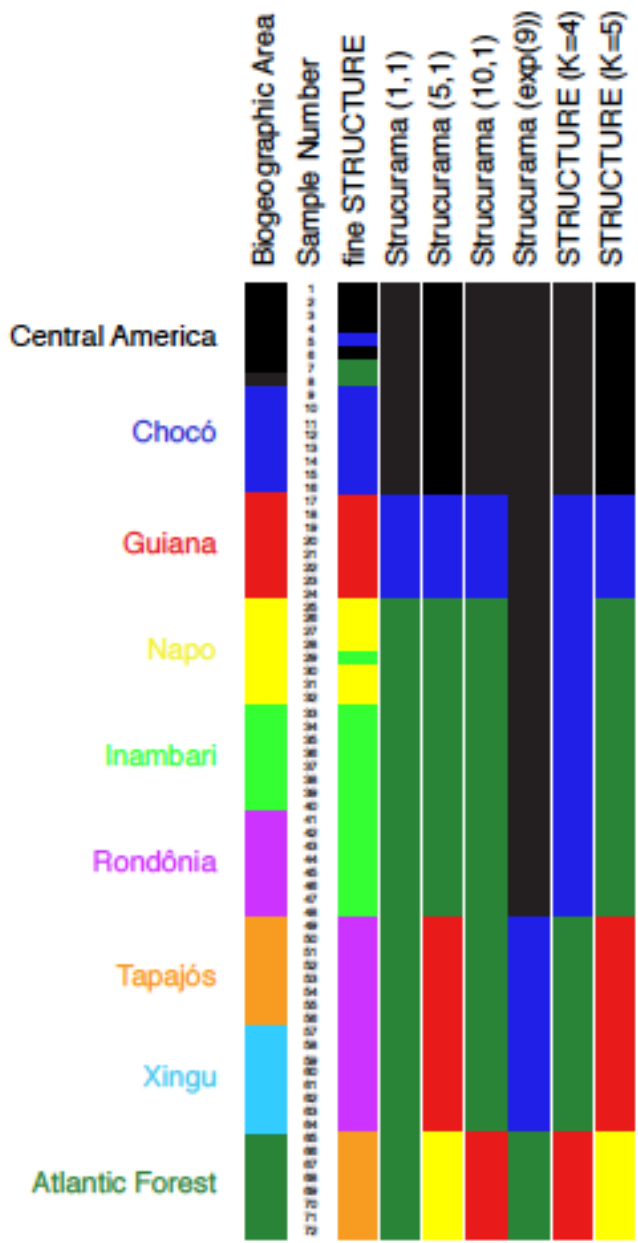

**Supp. Fig. 1.** Numbers of populations and population assignments from a subset of the clustering analyses conducted in fineSTRUCTURE, STRUCTURE, and Structurama. Colors in the population assignment columns distinguish populations, but are not necessarily related between columns, nor do they refer to biogeographic areas.

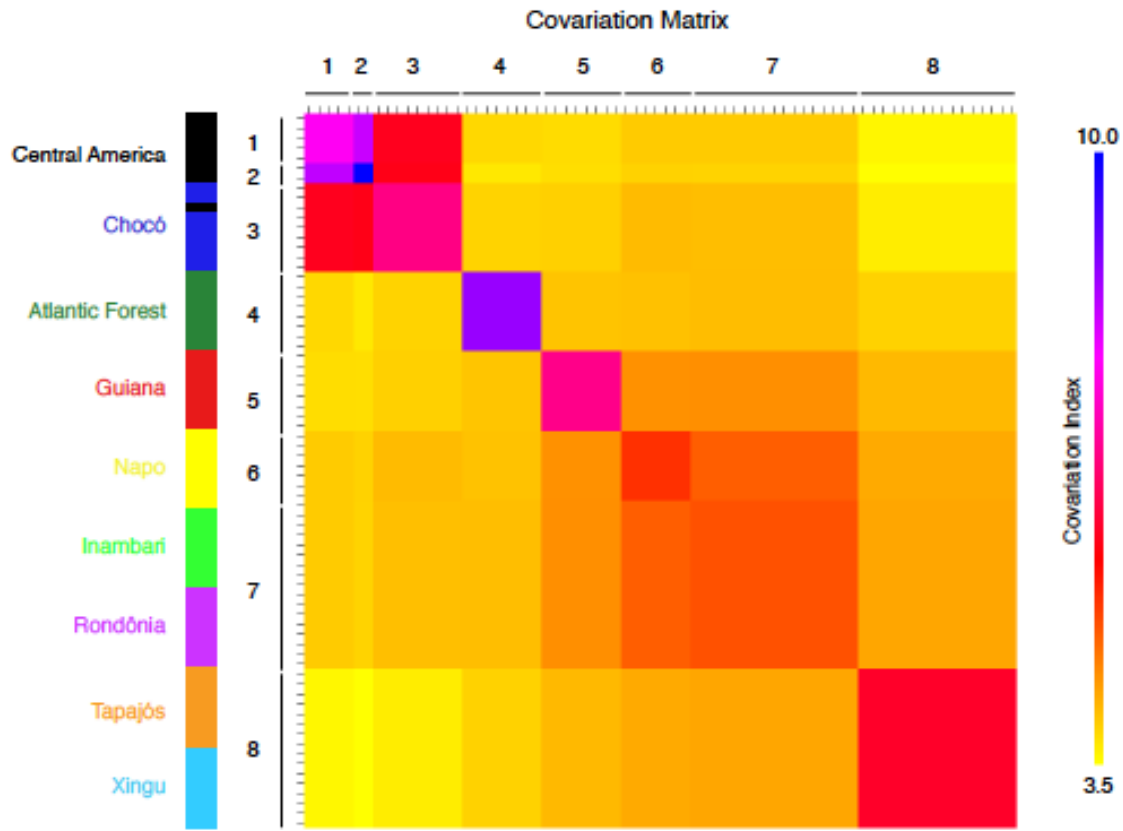

**Supp. Fig. 2.** The covariance matrix from the fineSTRUCTURE analysis showing populations identified by fineSTRUCTURE and the geographic area associated with each individual. Higher values in the covariation index correspond to greater similarity between individuals.

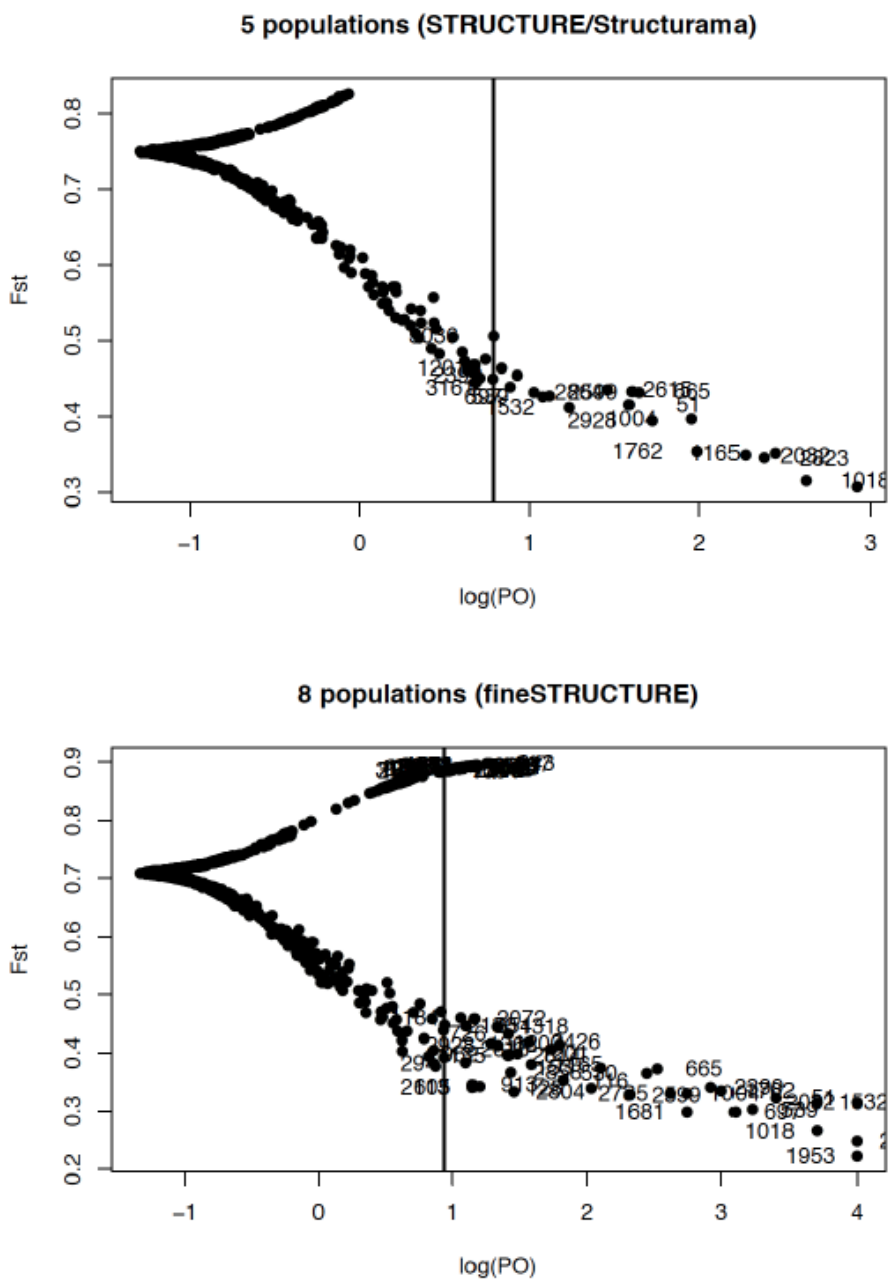

**Supp. Fig. 3.** Plots of $F_{st}$ for all loci from the BayeScan outlier analysis of both the STRUCTURE/Structurama populations and fineSTRUCTURE populations show positive (fineSTRUCTURE) and negative (both analyses) outliers that exceed the posterior odds ratio set based on an expected false discovery rate of 0.05.

**Table 1.** Influence of barriers on genetic variation in X. minutus. Partial Mantel test r-statistics measure the relationship between pairwise kinship estimates and whether the two individuals are on the same or opposite sides of a barrier, controlling for geographic distance (lower r-statistics indicate a stronger relationship). The BEDASSLE $\alpha_E/\alpha_D$ ratio measures the relative impact of barriers versus geographic distance on genetic similarity (higher values indicate a stronger relationship).

| Dataset | partial Mantel test r-statistic (SE) | BEDASSLE $\alpha E/\alpha D$ ratio (credible interval) |
|---|---|---|
| Isolation by Barriers | | |
| All barriers | -0.647 (-0.676, -0.612)* | 0.416 (0.276, 0.588) |
| Isthmus of Panama | -0.716 (-0.809, -0.646)* | 0.0773 (0.0619, 0.0975) |
| Andes Mountains | -0.737 (-0.798, -0.620)* | 137 (22.3, 466) |
| Rio Negro | -0.797 (-0.843, -0.736)* | 62.2 (21.5, 129) |
| Rio Solimões | -0.519 (-0.830, -0.359)* | 0.125 (0.0781, 0.189) |
| Rio Madeira | -0.469 (-0.661, -0.357)* | 0.0168 (0.00905, 0.0271) |
| Rio Tapajós | -0.844 (-0.924, -0.800)* | 99.0 (35.3, 324) |
| Rio Xingu | -0.276 (-0.410, -0.180)* | 0.0296 (0.0150, 0.0682) |
| Cerrado Belt | -0.531 (-0.712, -0.421)* | 136 (10.8, 8,060) |

*P<0.001*

**Table 2.** Theta (θ) and population growth rate (g) estimates from LAMARC for each

STRUCTURE/Structurama and fineSTRUCTURE population.

| Population | θ (95% CI) | g (95% CI) |
|---|---|---|
| STRUCTURE/Structurama | | |
| 1 | 5.2 (2.9, 9.2) | 64.4 (48.8, 75.3) |
| 2 | 8.4 (2.2, 9.8) | 70.6 (52.7, 94.3) |
| 3 | 9.9 (6.9, 10.0) | 55.7 (47.5, 63.1) |
| 4 | 8.1 (3.7, 9.8) | 120.6 (94.8, 133.8) |
| 5 | 1.0 (0.4, 5.2) | 174.3 (112.0, 241.3) |
| fineSTRUCTURE | | |
| 1 | 8.7 (0.4, 9.8) | 91.9 (-170.2, 208.4) |
| 2 | 5.7 (0.5, 9.5) | 87.5 (57.7, 212.1) |
| 3 | 5.2 (1.9, 9.5) | 80.4 (54.5, 100.0) |
| 4 | 9.5 (2.9, 9.9) | 96.7 (68.2, 107.5) |
| 5 | 2.6 (1.1, 5.7) | 42.0 (32.7, 57.4) |
| 6 | 9.9 (6.8, 10.0) | 66.5 (57.0, 76.9) |
| 7 | 8.1 (3.3, 9.8) | 119.9 (90.7, 134.3) |
| 8 | 1.1 (0.4, 3.9) | 204.3 (120.6, 258.9) |

**Table 3.** LAMARC estimates of migration rate (M) between populations for both the STRUCTURE/Structurama populations and fineSTRUCTURE populations.

| Populations | M (95% CI) |
| --- | --- |
| STRUCTURE/Structurama | |
| 1 -> 3 | 0.0 (0.0, 0.2) |
| 3 -> 1 | 0.8 (0.0, 2.6) |
| 2 -> 3 | 3.3 (0.9, 7.2) |
| 3 -> 2 | 3.8 (0.4, 10.6) |
| 3 -> 4 | 0.9 (0.0, 3.5) |
| 4 -> 3 | 0.4 (0.0, 1.5) |
| 4 -> 5 | 2.0 (0.1, 8.7) |
| 5 -> 4 | 0.0 (0.0, 0.6) |
| fineSTRUCTURE | |
| 1 -> 2 | 31.6 (2.5, 92.9) |
| 2 -> 1 | 90.7 (12.5, 99.7) |
| 1 -> 3 | 2.6 (0.0, 9.6) |
| 3 -> 1 | 2.5 (0.1, 37.9) |
| 3 -> 5 | 0.0 (0.0, 0.6) |
| 5 -> 3 | 1.2 (0.0, 4.2) |
| 4 -> 5 | 0.0 (0.0, 0.6) |
| 5 -> 4 | 1.2 (0.0, 4.9) |
| 5 -> 6 | 4.3 (2.0, 8.6) |
| 6 -> 5 | 0.3 (0.0, 1.8) |
| 6 -> 7 | 1.9 (0.2, 5.2) |
| 7 -> 6 | 0.0 (0.0, 0.3) |
| 7 -> 8 | 4.3 (0.1, 12.3) |
| 8 -> 7 | 0.0 (0.0, 0.5) |

**Supp. Table 1.** Sample information for all individuals used in this study. The number column refers to sample numbers referenced elsewhere in the paper. Museum abbreviations correspond to: (ANSP) Academy of Natural Sciences of Drexel University, Philadelphia, USA; (CUMV) Cornell University Museum of Vertebrates, Ithaca, NY, USA; (KU) Kansas University Museum of Natural History, Lawrence, KS, USA; (LSUMZ) Louisiana State University Museum of Natural Science, Baton Rouge, LA, USA; (MZFC) Museo de Zoología "Alfonso L. Herrera" de la Facultad de Ciencias, Universidad Nacional Autónoma de México, DF, México; (MBM) Marjorie Barrick Museum, University of Nevada - Las Vegas, NV, USA now housed at the Burke Museum, University of Washington, Seattle, USA; (MPEG) Museu Paraense Emílio Goeldi, Belém, Brasil; (MZUSP) Museu de Zoologia da Universidade de São Paulo, Brasil; and (USNM) National Museum of Natural History, Smithsonian Institution, Washington, DC.

| # | Museum | Tissue Number | Biog. Area | Subspecies | Country | State | Locality | Lat. | Long. |
|---|---|---|---|---|---|---|---|---|---|
| 1 | LSUMZ | 35767 | Central America | *X. m. ridgwayi* | Costa Rica | Cartago | 11 km SW Pejibaye | 9.7833 | -83.7500 |
| 2 | LSUMZ | 60935 | Central America | *X. m. mexicanus* | Honduras | Cortés | Cerro Azul Meamber National Park, Los Pinos | 14.8728 | -87.9050 |
| 3 | USNM | 1283 | Central America | *X. m. ridgwayi* | Panama | Bocas Del Toro | Valiente Peninsula, Punta Alegre, N. Bahia Azul | 9.0215 | -81.7620 |
| 4 | CUMV | 50919 | Central America | *X. m. ridgwayi* | Panama | Chiriquí | Burica Peninsula, 100-160m | 8.0333 | -82.8667 |
| 5 | CUMV | 50738 | Central America | *X. m. ridgwayi* | Panama | Coclé | El Cope National Park | 8.6698 | -80.5930 |
| 6 | KU | 2044 | Central America | *X. m. mexicanus* | Mexico | Campeche | Calakmul, El Arroyo, 6 km S Silvituc | 18.5928 | -90.2561 |
| 7 | MZFC | 51 | Central America | *X. m. mexicanus* | Mexico | Chiapas | N portion of La Omega, Monumento Natural Yaxchilan | 16.9017 | -90.9733 |
| 8 | MZFC | 238 | Central America | *X. m. mexicanus* | Mexico | Oaxaca | 20 km NE Chalchijapa | 17.0667 | -94.5833 |
| 9 | ANSP | 2227 | Chocó | *X. m. littoralis* | Ecuador | Esmeraldas | 20 km NNW Alto Tambo | 1.0300 | -78.5800 |
| 10 | ANSP | 2315 | Chocó | *X. m. littoralis* | Ecuador | Esmeraldas | 20 km NNW Alto Tambo | 1.0300 | -78.5800 |
| 11 | LSUMZ | 11948 | Chocó | *X. m. littoralis* | Ecuador | Esmeraldas | El Placer | 0.8667 | -78.5500 |
| 12 | LSUMZ | 28753 | Chocó | *X. m. ridgwayi* | Panama | Colón | Road S-9 W off Gatun-Escobal Road (S-10), ca. 6 Kilometers SW Gatun | 9.2800 | -79.7100 |
| 13 | LSUMZ | 2209 | Chocó | *X. m. littoralis* | Panama | Darién | Cana on E slope Cerro Pirré | 7.7560 | -77.6840 |
| 14 | LSUMZ | 26932 | Chocó | *X. m. ridgwayi* | Panama | Panamá | Old Gamboa Road, 5 km NW Paraiso | 9.0583 | -79.6508 |
| 15 | UWBM | jmd270 | Chocó | *X. m. ridgwayi* | Panama | Panamá | Chagres National Park (old boyscout camp) | 9.2500 | -79.5830 |
| 16 | UWBM | gms1842 | Chocó | *X. m. ridgwayi* | Panama | Panamá | 20 km ESE Canita, Lago Bayano | 9.1532 | -78.6929 |
| 17 | USNM | 5132 | Guiana | *X. m. ruficaudus* | Guyana | Essequibo | Waruma River, E bank, ca. 15 river km S Kako River | 5.5000 | -60.7833 |
| 18 | USNM | 10887 | Guiana | *X. m. ruficaudus* | Guyana | Northwest | North Side Acari Mountains | 1.3833 | -58.9333 |
| 19 | USNM | 9333 | Guiana | *X. m. ruficaudus* | Guyana | Northwest | Baramita | 7.3667 | -60.4833 |
| 20 | KU | 3879 | Guiana | *X. m. ruficaudus* | Guyana | Cuyuni-Mazaruni | N slope Mount Roraima | 5.2167 | -60.7500 |
| 21 | LSUMZ | 45809 | Guiana | *X. m. ruficaudus* | Suriname | Sipaliwini | Lely Gegberte | 4.2744 | -54.7391 |
| 22 | AMNH | 12699 | Guiana | *X. m. ruficaudus* | Venezuela | Amazonas | Rio Baria, Cerro de la Neblina base camp | 0.8342 | -66.1667 |
| 23 | AMNH | 8845 | Guiana | *X. m. ruficaudus* | Venezuela | Amazonas | Mrakapiwie | 1.8954 | -65.0456 |
| 24 | AMNH | 11942 | Guiana | *X. m. ruficaudus* | Venezuela | Bolivar | 40 kM E Tumaremo on road to Bochinche | 7.3833 | -61.2167 |
| 25 | FMNH | 456908 | Napo | *X. m. obsoletus* | Brazil | Amazonas | Japurá, Rio Mapari | -2.0497 | -67.2631 |
| 26 | FMNH | 456909 | Napo | *X. m. obsoletus* | Brazil | Amazonas | Japurá, Rio Mapari | -2.0497 | -67.2631 |
| 27 | MPEG | JAP 231 | Napo | *X. m. obsoletus* | Brazil | Amazonas | Japurá, Rio Mapari | -2.0421 | -67.2879 |
| 28 | MPEG | JAP 299 | Napo | *X. m. obsoletus* | Brazil | Amazonas | Japurá, Rio Mapari | -2.0421 | -67.2879 |
| 29 | ANSP | 1484 | Napo | *X. m. obsoletus* | Ecuador | Morona-Santiago | Santiago | -3.4000 | -78.5500 |
| 30 | LSUMZ | 4244 | Napo | *X. m. obsoletus* | Peru | Loreto | Lower Rio Napo, E bank Rio Yanayacu, ca. 90 km N Iquitos | -2.8200 | -73.2738 |
| 31 | LSUMZ | 6862 | Napo | *X. m. obsoletus* | Peru | Loreto | 5 km N Amazon River, 85 km NE Iquitos | -3.4167 | -72.5833 |
| 32 | LSUMZ | 7127 | Napo | *X. m. obsoletus* | Peru | Loreto | 5 km N Amazon River, 85 km NE Iquitos | -3.4167 | -72.5833 |
| 33 | LSUMZ | 9026 | Inambari | *X. m. obsoletus* | Bolivia | Pando | Nicolás Suarez, 12 km by road S Cobija, 8 km W on road to Mucden | -11.4703 | -68.7786 |
| 34 | MPEG | ESEC 225 | Inambari | *X. m. obsoletus* | Brazil | Acre | ESEC Rio Acre, ca. 78 kM W Assis, Brasil | -11.0568 | -70.2713 |
| 35 | MPEG | UFAC 1858 | Inambari | *X. m. obsoletus* | Brazil | Acre | Feijó, Rio Envira, Novo Porto, Foz do Ig. Paraná do Ouro | -8.4599 | -70.5564 |
| 36 | MPEG | UFAC 815 | Inambari | *X. m. obsoletus* | Brazil | Acre | Rio Branco, Transacreana (AC-090) km 70, Ramal Jarinal km 11 | -9.9006 | -68.4756 |
| 37 | MPEG | UFAC 879 | Inambari | *X. m. obsoletus* | Brazil | Acre | Rio Branco, Transacreana (AC-090) km 70, Ramal Jarinal km 11 | -9.9006 | -68.4756 |
| 38 | MPEG | PUC 131 | Inambari | *X. m. obsoletus* | Brazil | Amazonas | Tefé, Base Petrobras/Urucu, Papagaio | -4.8500 | -65.0667 |
| 39 | KU | 18530 | Inambari | *X. m. obsoletus* | Peru | Cusco | ca. Alto Manguriari | -12.5655 | -73.0878 |
| 40 | FMNH | 433364 | Inambari | *X. m. obsoletus* | Peru | Cusco | Consuelo, 15.9 km SW Pilcopata | -13.0167 | -71.4833 |
| 41 | FMNH | 391109 | Rondónia | *X. m. obsoletus* | Bolivia | Beni | Hacienda Los Angeles, 10 km E Riberalta | -11.0092 | -65.9952 |
| 42 | LSUMZ | 14752 | Rondónia | *X. m. obsoletus* | Bolivia | Santa Cruz | Serrania de Huanchaca, 25km SE Calorata Arco Iris | -14.4867 | -60.6753 |
| 43 | LSUMZ | 15114 | Rondónia | *X. m. obsoletus* | Bolivia | Santa Cruz | Velasco, 13 km SW Piso Firme | -13.7700 | -61.9500 |
| 44 | LSUMZ | 18175 | Rondónia | *X. m. obsoletus* | Bolivia | Santa Cruz | Velasco, Parque Nacional Noel Kempff Mercado 86 km ESE Florida | -14.8333 | -60.4167 |
| 45 | LSUMZ | 18534 | Rondónia | *X. m. obsoletus* | Bolivia | Santa Cruz | Velasco, Parque Nacional Noel Kempff Mercado 60 km ESE of Florida | -14.8400 | -60.7300 |
| 46 | MPEG | FPR 040 | Rondónia | *X. m. obsoletus* | Brazil | Amazonas | Maués, Flona do Pau Rosa, Comunidade Fortaleza | -3.9461 | -58.4561 |
| 47 | MPEG | FPR 103 | Rondónia | *X. m. obsoletus* | Brazil | Amazonas | Maués, Flona do Pau Rosa, Comunidade Sta. Teresa | -3.4000 | -57.7000 |
| 48 | MPEG | MPDS 650 | Rondónia | *X. m. obsoletus* | Brazil | Amazonas | Município de Humaitá, T. Indígena Parintintin, Aldeia Pupunha, Castanhal | -7.4667 | -62.8167 |
| 49 | MPEG | DED 323 | Tapajós | *X. m. genibarbis* | Brazil | Mato Grosso | Município Nova Bandeirante, right bank Rio Juruena, Fazenda Vale Verde | -10.2519 | -58.2850 |
| 50 | FMNH | 392023 | Tapajós | *X. m. genibarbis* | Brazil | Mato Grosso | Municipio Alta Floresta, upper Rio Teles Pires-Rio Cristalino | -9.9040 | -55.8810 |
| 51 | MPEG | BR163-070 | Tapajós | *X. m. genibarbis* | Brazil | Pará | Altamira, 30 km SW Castelo dos Sonhos, Fazenda Jamanxin | -8.3894 | -55.3702 |
| 52 | MPEG | BR163-181 | Tapajós | *X. m. genibarbis* | Brazil | Pará | Itaituba, 7 km NW Moraes de Almeida | -6.2021 | -55.6882 |
| 53 | MPEG | FLJA 029 | Tapajós | *X. m. genibarbis* | Brazil | Pará | Novo Progresso, margem esquerda Rio Jamanxim | -4.7000 | -56.4500 |

| 54 | MPEG | PIME 017 | Tapajós | *X. m. genibarbis* | Brazil | Pará | Belterra, Flona do Tapajós, Br 163 km 117 | -2.6333 | -54.9500 |
|----|------|----------|---------|--------------------|--------|------|-------------------------------------------|---------|----------|
| 55 | MPEG | PIME 131 | Tapajós | *X. m. genibarbis* | Brazil | Pará | Placas, Assentamento Comunidade Fortaleza | -3.4729 | -54.5655 |
| 56 | MPEG | WM344 | Tapajós | *X. m. genibarbis* | Brazil | Pará | Belterra, Flona do Tapajós, Santarém/Cuiabá, BR 163 Km 117 | -3.3561 | -54.9492 |
| 57 | FMNH | 391347 | Xingu | *X. m. genibarbis* | Brazil | Pará | Serra dos Carajas | -6.0783 | -50.2468 |
| 58 | FMNH | 391348 | Xingu | *X. m. genibarbis* | Brazil | Pará | Serra dos Carajas | -6.0783 | -50.2468 |
| 59 | FMNH | 456904 | Xingu | *X. m. genibarbis* | Brazil | Pará | Portel, FLONA do Caxiuanã, Plot PPBIO | -1.9500 | -51.6000 |
| 60 | FMNH | 456905 | Xingu | *X. m. genibarbis* | Brazil | Pará | Portel, FLONA do Caxiuanã, Plot PPBIO | -1.9500 | -51.6000 |
| 61 | FMNH | 456906 | Xingu | *X. m. genibarbis* | Brazil | Pará | Portel, FLONA do Caxiuanã, Plot PPBIO | -1.9500 | -51.6000 |
| 62 | MPEG | FTA 023 | Xingu | *X. m. genibarbis* | Brazil | Pará | Carajás, FLONA Tapirapé-Aquiri | -2.9500 | -51.8667 |
| 63 | MPEG | MOP 048 | Xingu | *X. m. genibarbis* | Brazil | Pará | Ourilandia do Norte, Serra do Puma | -6.7490 | -51.0814 |
| 64 | MPEG | PPBIO 151 | Xingu | *X. m. genibarbis* | Brazil | Pará | Portel, FLONA do Caxiuanã, Plot PPBIO | -1.9500 | -51.6000 |
| 65 | MZUSP | 1667 | Atlantic | *X. m. minutus* | Brazil | São Paulo | Fazenda Barreiro Rico, São Paulo | -23.7114 | -47.4188 |
| 66 | MZUSP | 685 | Atlantic | *X. m. minutus* | Brazil | São Paulo | Piedade | -23.7114 | -47.4188 |
| 67 | MZUSP | 689 | Atlantic | *X. m. minutus* | Brazil | São Paulo | Piedade | -23.7114 | -47.4188 |
| 68 | KU | 255 | Atlantic | *X. m. minutus* | Paraguay | Caazapá | San Rafael National Park | -26.3796 | -55.6456 |
| 69 | KU | 293 | Atlantic | *X. m. minutus* | Paraguay | Caazapá | San Rafael National Park | -26.3796 | -55.6456 |
| 70 | KU | 342 | Atlantic | *X. m. minutus* | Paraguay | Caazapá | San Rafael National Park | -26.3796 | -55.6456 |
| 71 | KU | 373 | Atlantic | *X. m. minutus* | Paraguay | Caazapá | San Rafael National Park | -26.3796 | -55.6456 |
| 72 | LSUMZ | 25938 | Atlantic | *X. m. minutus* | Paraguay | Caazapá | Cord. de Caaguazu, 7.5 km E San Carlos | -26.1000 | -55.7667 |

**Supp. Table 2.** Options used in the UNEAK pipeline for data processing.

| Plug-in | Option | Value | Description |
|---|---|---|---|
| UMergeTaxaTagCountPlugin | -m | 200000000 | Maximum tag number in the merged TagCount file. Default: 60000000 |
| UmergeTaxaTagCountPlugin | -c | 5 | Minimum count of a tag must be present to be output. Default: 5 |
| UmergeTaxaTagCountPlugin | -t | | Merge identically named taxa or not. -t n = do not merge. Default: merge |
| UTagCountToTagPairPlugin | -e | 0.03 | Error tolerance rate in the network filter. Default: 0.03 |
| UMapInfoToHapMapPlugin | -mnMAF | 0.05 | Minimum minor allele frequency. Default: 0.05 |
| UMapInfoToHapMapPlugin | -mxMAF | 0.5 | Maximum minor allele frequency. Default: 0.5 |
| UMapInfoToHapMapPlugin | -mnC | 0 | Minimum call rate (proportion of taxa covered by at least one tag) |
| UMapInfoToHapMapPlugin | -mxC | 1 | Maximum call rate. Default: 1 (proportion of taxa covered by at least one tag) |

**Supp. Table 3.** Processing statistics from the UNEAK pipeline.

|  | Mean | Median | Standard Deviation |
| --- | --- | --- | --- |
| Individual (Taxa) Depth | 5.1253 | 5.0640 | 1.3068 |
| Site Depth | 4.9402 | 3.9251 | 4.4381 |
| Individual (Taxa) Missingness | 0.6776 | 0.6744 | 0.0528 |
| Site Missingness | 0.6776 | 0.8000 | 0.3003 |

**Supp. Table 4.** Results of aligning GBS loci to the Zebra Finch (*Taeniopygia guttata*) genome. Presented are the count of loci with the best-scoring blastn hit falling on each *T. guttata* chromosome for the 3,247 loci that mapped successfully.

| Zebra Finch Chromosome | Number of loci with highest-scoring blastn hit | Assembly Size (Mb) in Zebra Finch |
|---|---|---|
| 1 | 162 | 118.550 |
| 1A | 186 | 73.660 |
| 1B | 3 | 1.080 |
| 2 | 248 | 156.410 |
| 3 | 250 | 112.620 |
| 4 | 141 | 69.780 |
| 4A | 99 | 20.700 |
| 5 | 156 | 62.380 |
| 6 | 96 | 36.310 |
| 7 | 94 | 39.840 |
| 8 | 97 | 27.990 |
| 9 | 87 | 27.240 |
| 10 | 68 | 20.810 |
| 11 | 73 | 21.400 |
| 12 | 66 | 21.580 |
| 13 | 78 | 16.960 |
| 14 | 84 | 16.420 |
| 15 | 74 | 14.430 |
| 16 | 0 | 0.010 |
| 17 | 56 | 11.650 |
| 18 | 71 | 11.200 |
| 19 | 58 | 11.590 |
| 20 | 66 | 15.650 |
| 21 | 40 | 5.980 |
| 22 | 15 | 3.370 |
| 23 | 38 | 6.200 |
| 24 | 38 | 8.020 |
| 25 | 9 | 1.280 |
| 26 | 36 | 4.910 |
| 27 | 28 | 4.620 |
| 28 | 35 | 4.960 |
| LGE22 | 0 | 0.883 |
| LG2 | 0 | 0.110 |
| LG5 | 0 | 0.016 |
| Z | 142 | 72.860 |
| MT | 0 | 0.017 |
| Unknown | 553 | 174.340 |

**Supp. Table 5.** Mantel and partial Mantel test results. A dash (-) separates the variables being examined, while a comma (,) precedes the variable being controlled for in partial Mantel tests.

| Dataset | Test | r-statistic (95% CI) | p-value |
|---|---|---|---|
| **Isolation by Distance** | | | |
| All areas | Mantel (Geography - Fij) | -0.4964 (-0.5211, -0.4783) | 0.0001* |
| All areas | Partial Mantel (Geography - Fij, Barriers) | -0.3133 (-0.3461, -0.2860) | 0.0001* |
| Central America | Mantel (Geography - Fij) | -0.1225 (-0.4487, 0.1153) | 0.3485 |
| Chocó | Mantel (Geography - Fij) | -0.3425 (-0.6126, -0.0575) | 0.0605 |
| Guiana | Mantel (Geography - Fij) | -0.3588 (-0.5673, -0.1277) | 0.1769 |
| Napo | Mantel (Geography - Fij) | -0.4069 (-0.4741, -0.3612) | 0.0081* |
| Inambari | Mantel (Geography - Fij) | -0.2762 (-0.5604, 0.2833) | 0.3738 |
| Rondônia | Mantel (Geography - Fij) | -0.5859 (-0.8680, -0.3955) | 0.0072* |
| Tapajós | Mantel (Geography - Fij) | -0.1646 (-0.4353, 0.0434) | 0.4317 |
| Xingu | Mantel (Geography - Fij) | -0.2824 (-0.4095, -0.02415) | 0.1105 |
| Atlantic Forest | Mantel (Geography - Fij) | -0.5816 (-0.8176, -0.3454) | 0.0032* |
| **Isolation by Barriers** | | | |
| All barriers | Partial Mantel (Barrier - Fij, Geography) | -0.6467 (-0.6762, -0.6123) | 0.0001* |
| Isthmus of Panama | Partial Mantel (Barrier - Fij, Geography) | -0.7158 (-0.8085, -0.6461) | 0.0001* |
| Andes Mountains | Partial Mantel (Barrier - Fij, Geography) | -0.7373 (-0.7978, -0.6203) | 0.0001* |
| Rio Negro | Partial Mantel (Barrier - Fij, Geography) | -0.7969 (-0.8432, -0.7362) | 0.0001* |
| Rio Solimões | Partial Mantel (Barrier - Fij, Geography) | -0.5187 (-0.8303, -0.3586) | 0.0001* |
| Rio Madeira | Partial Mantel (Barrier - Fij, Geography) | -0.4689 (-0.6611, -0.3568) | 0.0015* |
| Rio Tapajós | Partial Mantel (Barrier - Fij, Geography) | -0.8435 (-0.9236, -0.7997) | 0.0004* |
| Rio Xingu | Partial Mantel (Barrier - Fij, Geography) | -0.2756 (-0.4101, -0.1796) | 0.0074* |
| Cerrado Belt | Partial Mantel (Barrier - Fij, Geography) | -0.5313 (-0.7121, -0.4212) | 0.0002* |
| **Population Validation** | | | |
| STRUCTURE/Structurama populations | Partial Mantel (Populations - Fij, Geography) | -0.7611 (-0.7937, -0.7282) | 0.0001* |

| fineSTRUCTURE populations | Partial Mantel (Populations - Fij, Geography) | -0.6709 (-0.7167, -0.6293) | 0.0001* |

* *P*<0.01

**Supp. Table 6.** Results from STRUCTURE runs.

| K | Pr(X\|K) | Pr(K) |
|---|----------|-------|
| 1 | -167422.6 | -166580.1 |
| 2 | -128202.2 | -126761.2 |
| 3 | -100979.9 | -99679.6 |
| 4 | -77017.2 | -75504 |
| 5 | -65265.1 | -63458.3 |
| 6 | -77045.9 | -75514.4 |
| 7 | -77065 | -75520 |
| 8 | -65323.2 | -63480.1 |
| 9 | -65352 | -63487.1 |
| 10 | -65366.5 | -63494 |
| 11 | -77116.3 | -75539.6 |
| 12 | -65402.3 | -63506.8 |
| 13 | -65423.1 | -63513.7 |
| 14 | -65440.2 | -63520.6 |
| 15 | -65452.1 | -63526.5 |